\documentclass[article]{aa}
\usepackage{natbib}
\usepackage{graphicx}
\usepackage{txfonts}
\usepackage[colorlinks=true,citecolor=blue]{hyperref}
\usepackage{color}
\usepackage{xspace}

\bibpunct{(}{)}{;}{a}{}{,}

\newcommand{\MJup}{M$_{\mathrm{Jup}}$\xspace}

\newcommand{\mic}{$\mu$m\xspace}
\newcommand{\as}{\hbox{$^{\prime\prime}$}\xspace}

\begin{document}
\title{Performance of the VLT Planet Finder SPHERE I.}
   \subtitle{Photometry and astrometry precision with IRDIS and IFS in laboratory}

   \author{A. Zurlo\inst{1,2}, A. Vigan\inst{1}, D. Mesa\inst{2}, R. Gratton\inst{2}, C. Moutou\inst{1}, M. Langlois\inst{3}, R.U. Claudi\inst{2}, L. Pueyo\inst{4}, A. Boccaletti\inst{5},\\A. Baruffolo\inst{2}, J.-L. Beuzit\inst{6}, A. Costille\inst{1,6}, S. Desidera\inst{2},  K. Dohlen\inst{1}, M. Feldt\inst{7}, T. Fusco\inst{1}, T. Henning\inst{7}, M. Kasper\inst{8}, P. Martinez\inst{9}, O. Moeller-Nilsson\inst{7}, D. Mouillet\inst{6}, A. Pavlov\inst{7}, P. Puget\inst{6}, J.-F. Sauvage\inst{6,10}, M. Turatto\inst{2}, S. Udry\inst{11}, F. Vakili\inst{12}, R. Waters\inst{13}, R.F. Wildi\inst{11}  }

\institute{\inst{1} Aix Marseille Universit\'e, CNRS, LAM (Laboratoire d'Astrophysique de Marseille) UMR 7326, 13388, Marseille, France\\
\inst{2} INAF-Osservatorio Astronomico di Padova, Vicolo dell'Osservatorio 5, 35122, Padova, Italy \\
\inst{3}CRAL, UMR 5574, CNRS, Universit\'e Lyon 1, 9 avenue Charles Andr\'e, 69561 Saint Genis Laval Cedex, France\\
\inst{4}Johns Hopkins University, Department of Physics and Astronomy, 366 Bloomberg Center 3400 N. Charles Street, Baltimore, MD 21218, USA\\
\inst{5}LESIA, Observatoire de Paris, CNRS, University Pierre et Marie Curie Paris 6 and University Denis Diderot 
         Paris 7, 5 place Jules Janssen, F-92195 Meudon, France\\ 
%\inst{5}Space Telescope Science Institute, 3700 San Martin Drive, Baltimore, MD 21218, USA
\inst{6}UJF-Grenoble 1 / CNRS-INSU, Institut de Plan\'etologie et d'Astrophysique de Grenoble (IPAG) UMR 5274, Grenoble, F-38041, France\\
\inst{7}Max-Planck-Institut f\"ur Astronomie, K\"onigstuhl 17, 69117 Heidelberg, Germany\\
 \inst{8}European Southern Observatory, Karl-Schwarzschild-Strasse 2, D-85748 Garching, Germany\\
\inst{9}Laboratoire Lagrange, UMR7293, Universit\'e de Nice Sophia-Antipolis, CNRS, Observatoire de la Cote d’Azur, Bd. de l’Observatoire, 06304 Nice, France\\
\inst{10}ONERA - The French Aerospace Lab BP72 - 29 avenue de la Division Leclerc FR-92322 CHATILLON CEDEX\\
\inst{11}Observatoire de Gen\`eve, University of Geneva, 51 Chemin des Maillettes, 1290, Versoix, Switzerland\\
\inst{12}ETH Zurich, Institute for Astronomy, Wolfgang-Pauli-Strasse 27, 8093 Zurich, Switzerland\\
\inst{13}Sterrenkundig Instituut Anton Pannekoek, University of Amsterdam, Science Park 904, 1098 Amsterdam, The Netherlands
}

   \date{Submitted/Accepted}

% \abstract{}{}{}{}{}
% 5 {} token are mandatory

  \abstract
   {The new planet finder for the Very Large Telescope (VLT), the Spectro-Polarimetric High-contrast Exoplanet REsearch (SPHERE), just had its first light in Paranal. A dedicated instrument for the direct detection of planets, SPHERE, is composed of a polametric camera in visible light, the Zurich IMager POLarimeter (ZIMPOL), and two near-infrared sub-systems: the Infra-Red Dual-beam Imager and Spectrograph (IRDIS), a multi-purpose camera for imaging, polarimetry, and long-slit spectroscopy, and the integral field spectrograph (IFS), an integral field spectrograph. }
   {We present the results obtained from the analysis of data taken during the laboratory integration and validation phase, after the injection of synthetic planets. Since no continuous field rotation could be performed in the laboratory, this analysis presents results obtained using reduction techniques that do not use the angular differential imaging (ADI) technique. }
   {To perform the simulations, we used the instrumental point spread function (PSF) and model spectra of L and T-type objects scaled in contrast with respect to the host star.  We evaluated the expected error in astrometry and photometry as a function of the signal to noise of companions, after spectral differential imaging (SDI) reduction for IRDIS and spectral deconvolution (SD) or principal component analysis (PCA) data reductions for IFS.  }
   {We deduced from our analysis, for example, that $\beta$\,Picb, a 12~Myr old planet of $\sim$10~\MJup and semi-major axis of 9--10 AU, would be detected with IRDIS with a photometric error of 0.16~mag and with a relative astrometric position error of 1.1~mas. With IFS, we could retrieve a spectrum with error bars of about 0.15~mag on each channel and astrometric relative position error of 0.6~mas.  For a fainter object such as HR\,8799d, a 13~\MJup planet at a distance of 27~AU, IRDIS could obtain a relative astrometric error of 3~mas.  }
{}

   \keywords{Instrumentation: high angular resolution, spectrographs, Methods: data analysis, Techniques: imaging spectroscopy, Stars: planetary systems}

\titlerunning{Photometry and astrometry with IRDIS and IFS}
\authorrunning{Zurlo et al.}
\maketitle
%
%________________________________________________________________
\section{Introduction}
Since the discovery of the first exoplanet around a main sequence star by the radial velocity (RV) technique~\citep{1995Natur.378..355M} more than 1000 of these objects have 
been detected up to now \citep[see, e.g., the Extrasolar Planets Encyclopaedia\footnote{http://exoplanet.eu/},][]{2011A&A...532A..79S}. The majority of these discoveries has been performed exploiting the RV and the transit techniques. These two techniques are biased toward planets in close orbits. For this reason, the mass vs. semi-major axis parameter space is not homogeneously covered, and at the moment, just 70 planetary mass objects have been found at a separation larger than 5~AU. Among these long period objects, 40 have been found using the direct imaging technique, and 23 using the RV technique. For the latter objects, the orbital parameters are not well constrained when they show only long-term trends rather than a full orbital period. Direct imaging is then a complementary technique to explore the outer separations of the exoplanetary systems. In particular, this technique allows us to study the regions beyond the snow line around young stars. Also, taking advantage of the intrinsic luminosity of young giant gaseous planets in the first phases of their evolution, we can infer their masses \citep[see, e.g., evolutionary models by][]{2000ApJ...542..464C,2002A&A...382..563B, 2003A&A...402..701B,2005AN....326..925F, 2006ApJ...640.1063B}. However, the mass determination is subject to unconstrained physics and unknown initial conditions at very young ages \citep{2007ApJ...655..541M,2012ApJ...745..174S} and large discrepancies on the derived mass are expected between the ``hot-start'' and ``warm-start'' models.
%In general planets at wide separations are rarer around late spectral types stars
%while they are more common around A spectral type stars~\citep{2012A&A...544A...9V, 2013ApJ...777..160B}.
%For this reason, this latter type of star is a preferable target for direct imaging survey.

Direct imaging provides insights on formation and migrations mechanisms for planetary systems. Moreover, this technique allows us to obtain photometric, spectroscopic, and astrometric measurements of the detected companions, and for this reason it is a fundamental technique to study the atmosphere of the known objects, their mass-luminosity function and their orbits \citep[see, e.g,][]{2013arXiv1305.7428R,2013arXiv1307.2886K, 2013A&A...549A..52E, 2013ApJ...776...15C} . However, direct detection of extrasolar planets is challenging for two reasons: (1) the large luminosity contrast with respect to the star, which is of the order of $10^{-6}$ for giant young planets with high intrinsic luminosity, and $10^{-8}$--$10^{-9}$ for old planets seen in reflected and intrinsic light; and (2) the small separation between the star and the planet, of the order of a few tenths of arcsec for planets at few AUs around stars at a distance up to 100 pc from the Sun. As a result, the light from the companion objects is completely overcome by the light of the host star. These difficulties explain why, at the moment, not a large number of exoplanets have been discovered through direct imaging. 
%Some recent results are reported e.g. 
%in \citet{2013ApJ...766L...1Q} that found a protoplanet candidate embedded in the circumplanetary disk of the 
%star HD100546, \citet{2013arXiv1305.7428R} that presented the discovery of a 4 -- 5~\MJup planet around HD95086
%and, finally a cold jovian planet of 4~\MJup that has been detected by \citet{2013arXiv1307.2886K} around the 
%star GJ504. This latter one is the oldest found through the direct imaging technique.
%Other very interesting cases discovered few years ago are the multi-planetary system around HR8799, composed by four different
%planets \citep[see, e.g,][]{2008Sci...322.1348M, 2012ApJ...753...14S, 2012DPS....4420007M, 2013A&A...549A..52E} 
%and the planet around the young star $\beta$ Pic \citep{2009A&A...493L..21L, 2010Sci...329...57L,2013A&A...555A.107B, 2013ApJ...776...15C}.

%All these examples challenge the formation model based on the core accretion, as they 
%orbit at wide separation from their host stars, while the model predicts a typical maximum separation of 10~AU.\\

Despite the challenging nature of this technique, several different surveys have been performed in the past decade to search for extrasolar planetary systems. The main objectives of these surveys were to populate the mass vs semi-major axis diagram at large separations, and to clarify the mechanisms of planetary formation. Among the most recent of these surveys we can cite the NACO-Large Program at the Very Large Telescope~\citep{2014arXiv1405.1559D, 2014arXiv1405.1560C}, the Strategic Exploration of Exoplanets and Disk with Subaru (SEEDS) at the Subaru Telescope~\citep{2013ApJ...773...73J, 2014arXiv1404.5335B}, the Near-Infrared Coronagraphic Imager (NICI) campaign at the Gemini South Telescope~\citep{2013ApJ...776....4N, 2013ApJ...777..160B}, the International Deep Planet Survey \citep{2012A&A...544A...9V} at the Very Large Telescope (VLT) and Gemini North, and a survey of young, nearby, and dusty stars to understand the formation of wide-orbit giant planets \citep{2013A&A...553A..60R} at VLT. 

A new generation of high-contrast imaging instruments specifically designed for direct imaging of extrasolar planets is now operational, such as the Project 1640 at the 5~m Palomar Telescope~\citep{Crepp11} which provides important science results \citep[see, e.g,][]{2013ApJ...768...24O}, or the Gemini Planet Finder \citep[GPI,][]{2014arXiv1403.7520M} at the Gemini South Telescope that just concluded its commissioning phase and already provides scientific results \citep{2014A&A...565L...4G}. A fourth instrument, the Coronagraphic High Angular Resolution Imaging Spectrograph~\citep[CHARIS,][]{PetLim13}, is expected to be operative at the Subaru Telescope at the end of 2015. In Europe, the Spectro-Polarimetric High-contrast Exo-planet REsearch instrument \citep[SPHERE,][]{2006Msngr.125...29B} just had its first light at VLT. 

In this paper, we present the foreseen performance for the characterization of extrasolar planets using the two instruments composing the near-infrared (NIR) arm of SPHERE, IRDIS and IFS. The outline of this paper is the following: in Sect.~\ref{sec:sphere}, we present the characteristics of SPHERE and the synergy between its two NIR focal instruments; in Sect.~\ref{sec:tests}, we present the tests and data acquisition made in laboratory, and in Sect.~\ref{sec:red}, we describe their reduction and analysis. In Sect.~\ref{sec:fp}, we describe in Sect.~\ref{sec:fp} the simulations performed with synthetic planets injection, and in Sect.~\ref{sec:results} we present the photometric and astrometric results performed using IRDIS and IFS. Finally, we present our conclusions in Sect.~\ref{sec:conclusions}.

%__________________________________________________________________

\section{SPHERE: a new planet hunter for the VLT}
\label{sec:sphere}

The new planet-finder instrument at VLT in Chile, SPHERE, just had its first light during the spring 2014. The principal goal of the instrument is to find and characterize giant, gaseous, long-period planets within the solar neighborhood. As formation models predict, young planets are hot and self-luminous, making their direct detection possible by masking the primary star with a coronagraph. Evolved systems can also be detected through their reflected, polarized, light. From the ground, Extreme Adaptive Optics (ExAO) systems are required to correct for the atmospheric turbulence at very high frequency.

SPHERE is composed of the following subsystems:
\begin{itemize}
  \item an ExAO system called SAXO (SPHERE AO for eXoplanet Observation) \citep{saxo} that produces a highly stabilized beam with a Strehl Ratio (SR) of more than 90\%;
  \item the Common Path and Infrastructure (CPI) that brings the telescope light to the three scientific modules. The CPI contains the deformable mirror (DM), relay optics such as toric mirrors \citep{2012A&A...538A.139H}, derotator, atmospheric dispersion compensators, and coronagraphs \citep{2011aoel.confE..75D};
  \item the three science subsystems working in the visible or the near-infrared. The Infra-Red Dual-beam Imager and Spectrograph \citep[IRDIS,][]{Do08} and the Integral Field Spectrograph \citep[IFS,][]{Cl08} can operate in parallel on infrared light in the range (0.95--2.32 $\mu$m), while the Zurich IMager POLarimeter \citep[ZIMPOL,][]{Th08} operates in the visible (0.60--0.90 $\mu$m). The three instruments cannot all work at the same time.
\end{itemize}

\noindent IRDIS allows for a wide range of observing modes, including:

\begin{itemize}
  \item dual-band imaging (DBI),
  \item long-slit spectroscopy (LSS),
  \item classical imaging (CI), and
  \item dual-polarimetric imaging (DPI).
\end{itemize}

There is a selection of 12 filters available for imaging, in broad-, medium- or narrow-band, and five different filter pairs are dedicated to the DBI mode \citep{Arthur}. The LSS mode with resolving powers of 50 and 500 is coupled to simple Lyot coronagraphy for the characterization of detected companions \citep{2008A&A...489.1345V}. The CI and DPI observing modes can also be used for the study of extended objects as disks. 

The subsystem IFS is an integral field spectrograph that allows for two spectral ranges \citep{Cl08}: 

\begin{itemize}
 \item $YJ$-mode (0.95 -- 1.35~$\mu$m) with a two-pixels resolving power of 
$\sim$ 50; 
 \item $YJH$-mode (0.95 -- 1.65~$\mu$m) with a two-pixels resolving power of 
$\sim$ 30; 
\end{itemize}

The IFS is a powerful instrument to explore the inner regions of planetary systems. Its 39 spectral channels enable a good speckle subtraction, allowing us to reach deep contrasts at small angular separation.

To allow for a parallel operation of the two NIR instruments, the light entering the telescope is split in two beams downstream of the coronagraphic mask, each instrument having its own set of Lyot stops. Two dichroic plates are available to allow for two different observing modes: IRDIFS mode, where IRDIS performs DBI observations in $H$ band, while IFS works in $YJ$-mode; and IRDIFS\_EXT mode, where IRDIS performs DBI in $K_S$ band, and IFS observes in $YJH$-mode \citep{2006Msngr.125...29B}. A general description of the two modes is in Table~\ref{t:tech}. These modes are designed for large surveys looking for young giant planets, and it is expected that the IRDIFS mode will be predominantly used. For this reason, we aim to present results obtained exploiting this particular observing mode. 

%The two sub-systems, when observing together, will work in pupil-stabilized mode on the sky to take advantage of the field rotation.

\begin{center}  
\begin{table}
\begin{minipage}{\columnwidth}
\caption[]{Technical description of the subsystems IRDIS and IFS the IRDIFS and IRDIFS\_EXT modes.} 
\label{t:tech}
\renewcommand{\footnoterule}{}  % to avoid a line before footnotes
\begin{tabular}{lll}
\hline
\hline
               &  IRDIS   & IFS  \\
\hline
FoV\footnote{Field of view} &    12\farcs5$\times$11\farcs0     & 1\farcs77$\times$1\farcs77 \\
Pixel scale            &  12.25 mas/px       &  7.30 mas/px\footnote{After the pipeline resampling} \\
Spectral range       &  0.95--2.32 $\mu$m        & 0.95--1.65 $\mu$m \\
Channels    &  2        &  39  \\
IRDIFS mode  &  $H2H3$\footnote{$\lambda_{H2}=1.59~\mu$m, $\lambda_{H3}=1.66~\mu$m}    &   $YJ$ \\
IRDIFS\_EXT   & $K1K2$\footnote{$\lambda_{K1}=2.10~\mu$m, $\lambda_{K2}=2.24~\mu$m}    &     $YJH$  \\
\hline
\end{tabular}
\end{minipage}
\end{table}
\end{center}

\section{Acquisition of the data in laboratory}
\label{sec:tests}

During the Assembly, Integration and Testing (AIT) phase, SPHERE was located at Institut de Plan\'etologie et d'Astrophysique de Grenoble (IPAG). During a period of more than a year, both IRDIS and IFS subsystems were tested in various configurations.

A telescope simulator (TSIM) with central obstruction and spiders matching that of the VLT was installed at the entrance of SPHERE to produce a realistic pupil and generate turbulent conditions. The turbulence is simulated by the use of rotating reflective phase screens, which allows us to tune the seeing and windspeed to recreate typical observing conditions in Paranal. One of the limitations of the TSIM is that the VLT pupil it contains is fixed (or rotated by 90 degrees steps) with respect to the instrument, which means that the SPHERE derotator has to remain fixed during the observations to avoid misalignment of the pupil with the Lyot stops. The main consequence is that angular differential imaging \citep[ADI;][]{2006ApJ...641..556M} could not be simulated properly in laboratory.

During IRDIFS tests, the calibration sequence of the instrument typically includes the acquisition of:
     
\begin{itemize}
  \item dark frames for IFS;
  \item flat-field frames for both instruments;
  \item a spectra position calibration for IFS;
  \item a wavelength calibration for IFS;
  \item an instrument flat for IFS that evaluates the response of all the lenslets of the integral field unit (IFU);
  \item IRDIS backgrounds taken in the same conditions (optical setup and exposure time) as the scientific images (off-axis point spread function (PSF) and coronagraphic images); and
\item two images of four symmetric replicae of the central PSF to calibrate the center of the star behind the coronograph.
\end{itemize}
 
The acquisition of scientific data generally follows the scheme described below. First, the fine centering of the star on the coronagraph is performed using an automated procedure that ensures both centering and focusing on the coronagraphic mask. This procedure is used for all coronagraphs available in SPHERE: three apodized Lyot coronagraphs \citep{2011ExA....30...39C} of different dimensions (145, 185, and 240 mas of diameter), two 4-quadrants phase mask coronagraphs \citep[see][]{Boc08}, as well as two classical Lyot coronagraphs.

%The scientific acquisition was performed with various configurations of the TSIM to explore different parameters of the simulated seeing conditions. 

In order to get a precise measurement of the star center behind the coronagraphic mask, a reference image with waffle spots is acquired immediately before or after the scientific sequence. Just one acquisition was needed as the tests confirmed that the system remained stable in the laboratory during the entire observation sequence. When observing on the sky, we will take one at the beginning and one at the end if proven necessary. The waffles are four replicae of the central PSF, placed in symmetrical positions around the center of the star. Waffles are introduced by the use of a small periodic offset on the deformable mirror in closed-loop operations, by an appropriate modification of the adaptive optics (AO) reference slopes \citep[as proposed in][]{2006ApJ...647..620S,2006ApJ...647..612M,2013aoel.confE..63L,2014arXiv1407.2308W}. The waffle pattern introduced has a spatial frequency of 10 cycles/aperture, and an amplitude of 74~nm, to create well illuminated but not saturated spots.

Before or after the coronagraphic sequence, an off-axis image of the PSF of the star is acquired by introducing an offset on its position. This off-axis PSF is acquired with the same configuration of the instrument to be able to perform flux calibration of the coronagraphic images. To avoid saturation we use a neutral density (ND) filter in the CPI to attenuate the flux of the beam. This filter attenuate the flux of the simulated star by a factor of $10^{3.5}$.

For the coronagraphic sequence of images, exposure times are adapted to the simulated conditions of seeing and stellar magnitude. To improve the flat-field accuracy from 0.5\% to 0.1\%, both instruments have the possibility of dithering their detector in the focal plane on a square grid of a few pixels (up to 10), by steps of one pixel. By this procedure, the scientifically useful signal falls on different physical pixels throughout the observing sequence, which results in averaging flat-field variations after the images are aligned and combined during the data analysis \citep{2008SPIE.7018E..29R}. For IRDIS, the improvement in flat field accuracy is expected to play a role in small separations ($\lesssim 0\farcs5$) for contrasts below a few $10^{-6}$, levels, which could not be reached in the laboratory because of the lack of ADI. Nonetheless, the dithering was used to be representative of a normal on-sky observing sequence.

In the case of IFS, dithering was thought to improve the instrument performance for contrasts better than $10^{-7}$ that cannot be reached in laboratory. However, previous tests have demonstrated that the use of dithering can degrade the final contrast obtained by IFS. This is given by the difficulties in defining in a precise way the spectra positions once dithering is implemented.

For IFS, the results discussed in this paper are obtained without dithering.
   
In general, the noise sources on the final raw images are:
\begin{itemize}
\item the readout noise, which depends on the readout mode chosen;
\item the thermal background, which varies with the temperature of the instrument and is homogeneous over the whole image;
\item the photon noise, which follows $\sqrt{N}$, where N is the number of detected photons; and
\item the speckle noise, which decreases with the separation from the star;
%\item the noise related to the calibrations (dark, flat-field, off-centered PSF, waffle centering, ...). 
\end{itemize}

To improve performance and limit as much as possible the instrumental noise level, a specific reduction process is required, as described in the following section.
  
\section{Data reduction and detection limits}
\label{sec:red}

We selected an observation sequence acquired in October 2013 at IPAG during the Preliminary Acceptance in Europe (PAE). The observing conditions of the instrument were as close as possible in terms of system calibration and performance when observing on sky. Moreover, we simulated a typical atmospheric condition of Paranal\footnote{http://www.eso.org/gen-fac/pubs/astclim/paranal/}, where the median value of the wind speed is $\sim 6m/s$ and the median seeing value is around $0.69\as$. This sequence includes calibrations and scientific datacubes as described in Sec.~\ref{sec:tests}. Information on the selected datasets are listed in Table~\ref{t:obs}.

\begin{table}
\begin{minipage}{\columnwidth}
\caption{Observation sequence characteristics used for the characterization analysis after the injections of the synthetic planets. The data were taken in the laboratory in IRDIFS-mode (see Table~\ref{t:tech}).} 
\label{t:obs}
\renewcommand{\footnoterule}{}  % to avoid a line before footnotes
\centering
\begin{tabular}{lcc}
\hline
\hline
                      &  IRDIS   & IFS  \\
\hline
Simulated seeing      & \multicolumn{2}{c}{0.85\as} \\
Simulated wind speed  & \multicolumn{2}{c}{5~m/s} \\
Coronagraph type      & \multicolumn{2}{c}{Apodized pupil Lyot coronagraph} \\
Coronagraph diameter  & \multicolumn{2}{c}{185~mas} \\
\hline
IRDIFS-mode           &  $H2H3$    &   $YJ$  \\
DIT\footnote{Detector Integration Time} &   1.6~s     & 2.0~s \\
NDIT\footnote{Number of frames per dithering position} & 20  & 50 \\
Total exposure time   &  512~s   &  100~s   \\
Image size            & 2048$\times$1024  &   2048$\times$2048  \\ 
Dithering             & 4$\times$4 &   None  \\
FoV rotation          & None & None  \\
ND\footnote{Neutral Density filter} & 0.0 & 0.0 \\
\hline
\end{tabular}
\end{minipage}
\end{table}

The sequence we performed simulates an observation of a bright star with $J=2.6$. The IRDIS raw images were prereduced performing background subtraction, bad-pixels correction, and flatfielding. The precise star center was measured on the waffle images acquired right before the coronagraphic sequence. This calibration measurement was used to deduce the coordinates of the center of the star for each frame taking the detector dithering positions into account.

After the preprocessing of each frame, spectral differential imaging \citep[SDI,][]{1999PASP..111..587R} was performed. The general principle of SDI is that two images of the star acquired simultaneously at close wavelengths can be subtracted to remove most of the stellar halo and speckle pattern. To ensure optimal rejection of the speckle noise, the images must be aligned, resampled on the same spatial scale, and scaled in intensity to account for any filter transmission and stellar flux variation. For this work, we implemented an optimized SDI procedure designed to minimize the speckle noise in the subtracted image between 0\farcs25 and 0\farcs65. The parameters that were optimized by the minimization routine are the amplitude scaling factor, and the differential shift between the images at the two wavelengths. The spatial scaling factor remain identical for all images, and equal to $\lambda_{H2}/\lambda_{H3}$. Because SPHERE images are at least Shannon-sampled, the spatial rescaling of the images acquired in H3 was performed accurately using a Fast Fourier Transform (FFT). The scaling consists of zero-padding the image both in direct space and in Fourier space in order to obtain the zoom factor of our choice \citep{2010SPIE.7735E..99V}.

\begin{figure}[h]
\begin{center}
\includegraphics[width=0.48\textwidth]{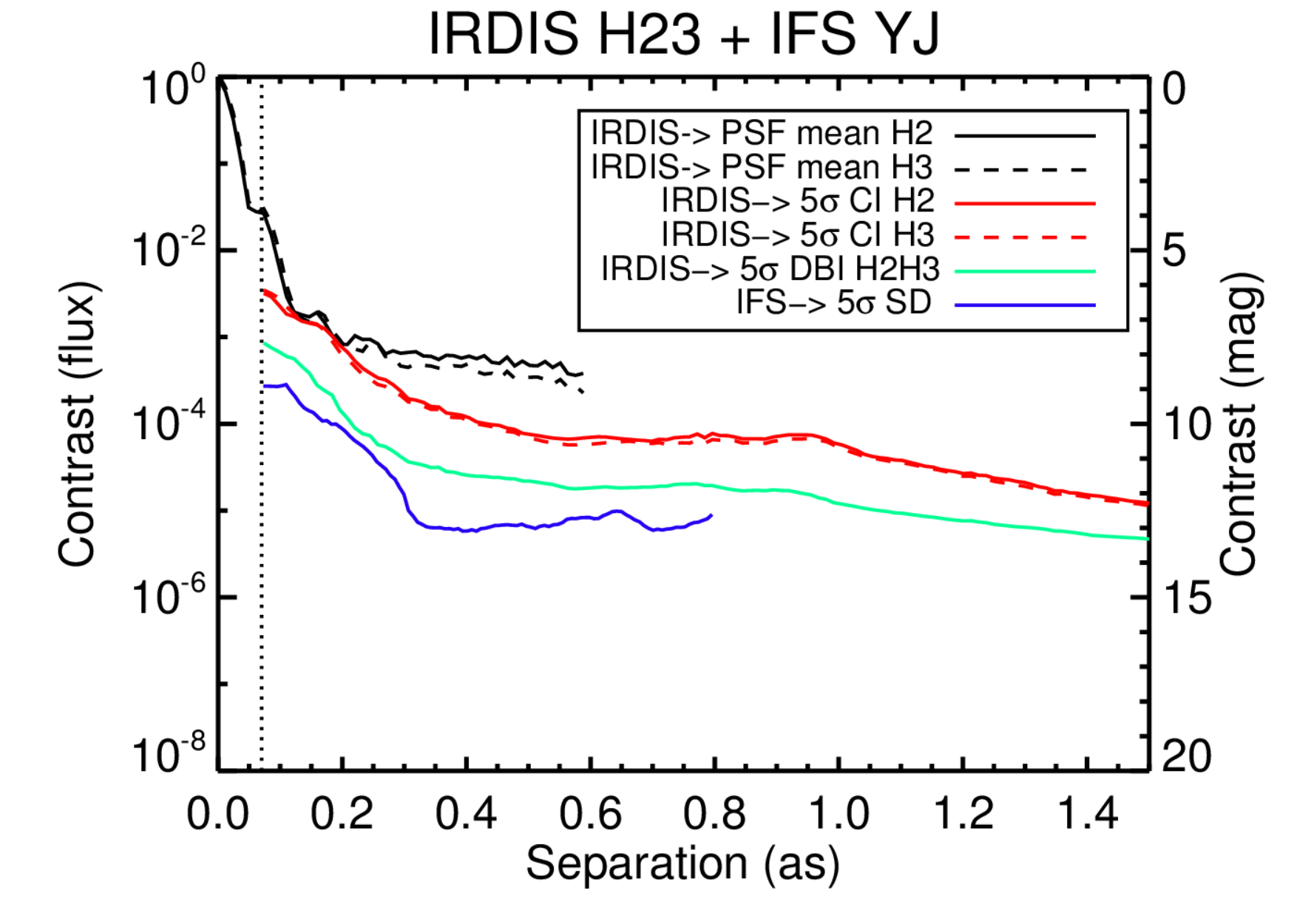}
\caption{5$\sigma$ residual noise levels, without ADI reduction, for the two science modules of SPHERE, IRDIS and IFS, in IRDIFS mode. The curves refer to the dataset described in Table~\ref{t:obs}, where we injected the synthetic planets. The vertical dotted line represents the radius of the coronagraph, i.e., the zone where the detectors are blind. The mean azimuthal profile of the off-axis PSF (black), the coronagraphic profile (red) are shown for the two IRDIS channels $H2$ (continuous line) and $H3$ (dashed line). The contrast results after the SDI reduction (green) and SD reduction for IFS (blue) are shown. }
\label{f:contrast}
\end{center}
\end{figure}

After averaging all SDI-subtracted frames, the final data product was used to calculate noise level as the residual standard deviation in the SDI image. The IRDIS DBI 5$\sigma$ noise level is plotted in Fig.~\ref{f:contrast}; red color represents the 5$\sigma$ noise level of individual channels, and green color the DBI curve. All the reductions of IRDIS images were done using custom IDL routines. The level of residual noise that we can obtain exploiting only SDI analysis are satisfactory, reaching a contrast of 2$\times10^{-5}$ at 0\farcs5 and 4.5$\times10^{-6}$ at 1\farcs5 from the primary at 5$\sigma$. Following the simulations presented in \citet{Arthur}, with the addition of the ADI technique, the contrasts will be of the order of 5$\times10^{-7}$ and $2.5\times10^{-7}$ at the same separations.

For IFS, the following calibrations were performed using the Data Reduction and Handling \citep[DRH,][]{Pa08} software, the official pipeline for the SPHERE instrument:

\begin{itemize}
\item dark subtraction; 
\item bad-pixels correction;
\item detector flat-fielding;
\item determination of all the spectra positions;
\item wavelength calibration; and
\item instrument flat correction to account for the different response of each lenslet.
\end{itemize}

After these calibrations, a scientific DRH recipe (module of the pipeline that performs a particular task) is applied to produce a three-dimensional datacube composed of 39 monochromatic images of 291$\times$291 pixels. 

On this datacube we used both the spectral deconvolution \citep[SD, see, e.g,][]{Sparks2002, Thatte2007} and the principal component analysis \citep[PCA, see, e.g,][]{2012MNRAS.427..948A, 2012ApJ...755L..28S,2013ApJ...768...24O} reductions methods. For more details, regarding the first method applied on SPHERE-IFS data we refer the reader to Mesa et al. (submitted).

Here we list the most important steps needed to implement the SD. First, each monochromatic frame in the datacube is rescaled according to its wavelength in such a way that the speckle pattern is at least ideally the same while the position of an eventual companion is different from frame to frame. After that a fit is performed along the rescaled datacube wavelength direction for each pixel. The fitting function is then subtracted from the rescaled datacube. The degree of the fitting function can be selected in such a way to maximize the speckle noise subtraction and, in the meantime, to avoid subtracting signal from the companion. Finally, each frame is rescaled back to its original dimension so that the companion is always in the same position. At separation less than the bifurcation radius ($\sim 0\farcs15$) \citep{Thatte2007} the spectrum is completely covered by an eventual companion that, for this reason, would be completely canceled so that the method is not effective at these small separations.
The SD method can provide good results on SPHERE data as the instrument meet some particular characteristics: the speckles do scale linearly and the chromaticity of the speckles can be completely described by a low order polynomial.

The method based on PCA, that we used for the characterization of the detected objects is presented in more details in Sect.~\ref{sub:klip}.

The procedure \citep[following][]{2005ApJ...625.1004M} that we have used to calculate the contrast is the following:
\begin{itemize}
\item the standard deviation into a box centered on each pixel and 
with a side corresponding to 1.5$\lambda /D$ is calculated for each pixel in the image;
\item a median of the standard deviation obtained as described in the previous point is made for all
the pixels at the same separation from the central star;
\item the contrast is defined by the ratio between the flux in the coronagraphic images and the flux of the central star. To calculate it we measured the flux of the off-axis PSF of the star after normalization for the value of the detector integration time (DIT) and the transmission of the neutral density (ND) filter used during the exposure, as mentioned in Sec.~\ref{sec:tests}. The same normalization for exposure time and ND filter was also applied to the coronagraphic images.
\end{itemize}

The contrast limit curve for IFS obtained exploiting the SD method described above is presented in Fig.~\ref{f:contrast}. In this case, we used a fitting function with degree of 1 because we found that in this manner we could obtain a good speckle subtraction simultaneously avoiding self-cancellation of the companion.

In this analysis the cancellation that the SD could cause is not taken into consideration and we refer the reader to Mesa et al. (submitted) for more details. The contrast obtained with the SD reduction is of the order of 5$\times10^{-6}$ at a separation of 0\farcs5 and 1$\times10^{-5}$ at a separation of 0\farcs8. With the addition of the ADI, we expect the contrast to reach 1.5$\times10^{-7}$ and 5$\times10^{-8}$ at the same separations, as resulted from the simulations presented in \citet{Dino}.

The contrast curves presented in Fig.~\ref{f:contrast} have been calculated on a short temporal sequence, so the speckles noise dominates over the other sources of noise described in Sect.~\ref{f:contrast}. We expect that longer sequences when observing on sky will improve the detection limits, as in the laboratory the TSIM reproduced identical speckles patterns with a given frequency.

\subsection{The KLIP method}
\label{sub:klip}

To improve our capability on the characterization of the planetary candidates that will be detected by IFS, we implemented a PCA method that performs the Karhunen--Lo{\`e}ve Image Projection (KLIP) algorithm, following the model of \citet{2012ApJ...755L..28S}, with improvements dedicated to the spectral extraction from IFS datacubes developed for Project 1640 data \citep{2014arXiv1409.6388P}. Our code was implemented in IDL language and is suitable to work with IFS datacubes. 

The KLIP method takes advantage of the multiple channels of the IFS to create the reference library for the basis of the Karhunen--Lo{\`e}ve matrix. The principle is that the signal of the planet after the spatial rescaling of the IFS datacube is in different positions with respect to the center of the image while the speckles pattern remains fixed. If we take a small portion of the image around the position of the planet in a specific spectral channel, it is possible to create a reference library using a characterization zone that is included in the projection of this portion on all the other channels that contain no signal from the planetary candidate, or at least or a very small quantity of this signal.

Using as a PCA library a set of portions of the frames that do not contain much astrophysical signal attenuates the typical flux losses induced by the SD technique. We used the forward modeling method presented in \citet{2012ApJ...755L..28S} to get our results in KLIP photometry, as we will show in Sect.~\ref{sub:phIFS}. On-sky observations are expected to produce even better results, as the construction of the PCA library will exploit the FoV rotation.

\section{Synthetic planets injection}
\label{sec:fp}

To estimate the errors on photometry and astrometry of future candidates, we injected synthetic planets in the set of data presented in Sect.~\ref{sec:red}. The synthetic planets consist of a small portion of the off-axis PSF acquired during the scientific sequence (see Sect.~\ref{sec:tests}). In this manner, the light diffracted by the spiders and the secondary ring of diffraction are taken into account. While for IRDIS it was possible to insert the planets directly into the raw data, for IFS we had to inject them in the datacube after the scientific recipe as described in Sect.~\ref{sec:red}. This is due to the fact that it is extremely difficult to inject the simulated objects into the IFS raw data, which is constituted by the thousands spectra created by all the IFS lenslets.

The flux in each spectral channel was calculated to reproduce L and T-type spectra. The libraries of field brown dwarfs used for the L-type spectra are taken from \citet{Testi2001}, while the T-type spectra are from \citet{2007AJ....134.1162L}, \citet{Burgasser2004} and \citet{Burgasser2006}. The flux ratio between the fluxes of the two IRDIS band ($H2$/$H3$) in the different spectral types ranged from a minimum value of 0.85 (L4-type, the flattest one) to a maximum of 7.46 (T8-type).

Five simulated planets were injected simultaneously at five different separations (0\farcs20, 0\farcs35, 0\farcs50, 0\farcs65, 0\farcs80) and position angles with respect to the star. The procedure was repeated 30 times with position angles rotated by steps of 12 degrees each time to improve the statistical significance of the results. The flux of the planets was scaled for five different contrast levels (10$^{-3}$, 3$\times$10$^{-4}$, 10$^{-4}$, 3$\times$10$^{-5}$ and 10$^{-5}$) with respect to the host star. The contrast was defined as the ratio between the integrated flux of the planet over that of the star, over the whole band covered by the two instruments (0.8--1.8 $\mu$m). This means that for a given contrast the flux collected by IFS is greater than for IRDIS, especially for T-type objects. Overall, the statistics of our results is based on a total of 750 injected planets (five planets each image, five contrasts, 30 different rotation angles) of the same spectral type. 

\begin{figure}[h]
\begin{center}
\includegraphics[width=0.48\textwidth]{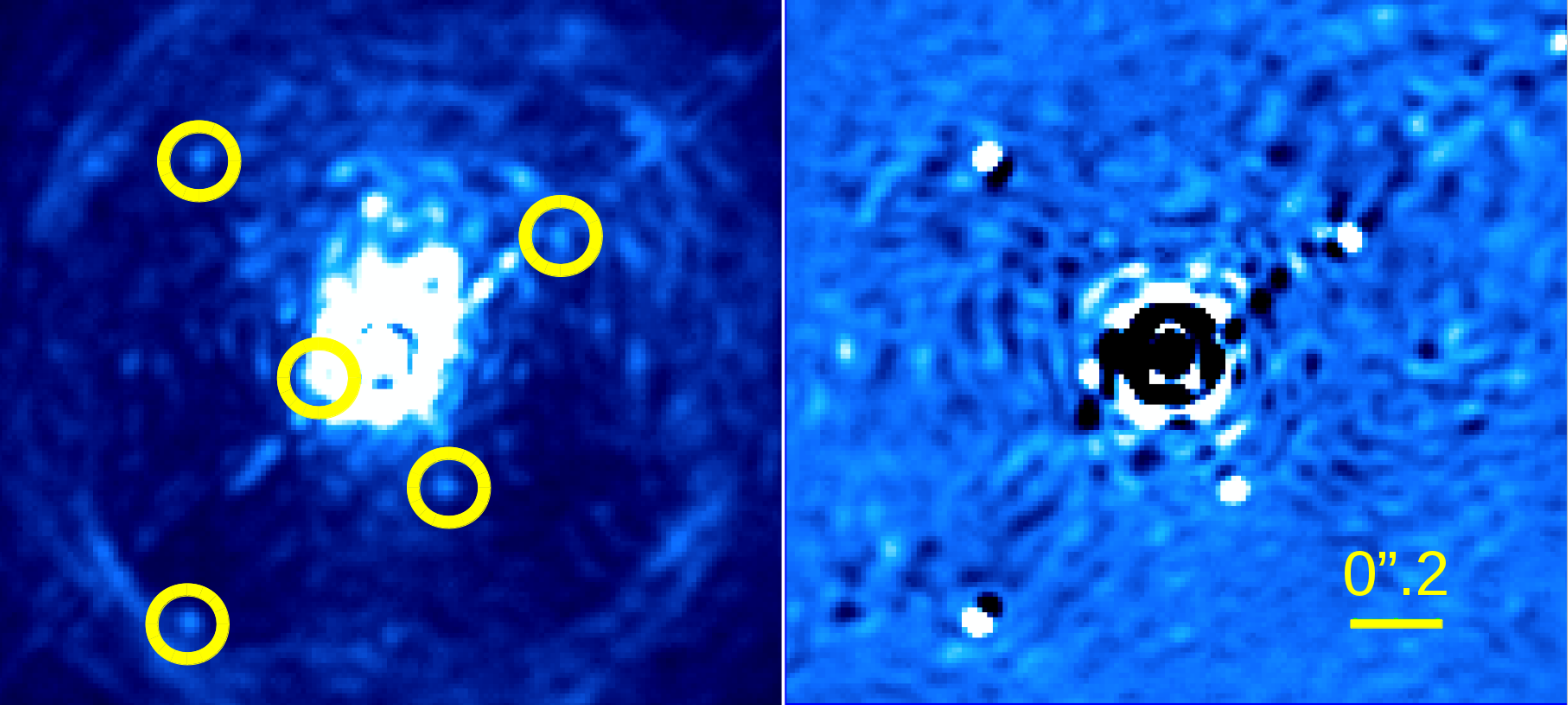}
\includegraphics[width=0.48\textwidth]{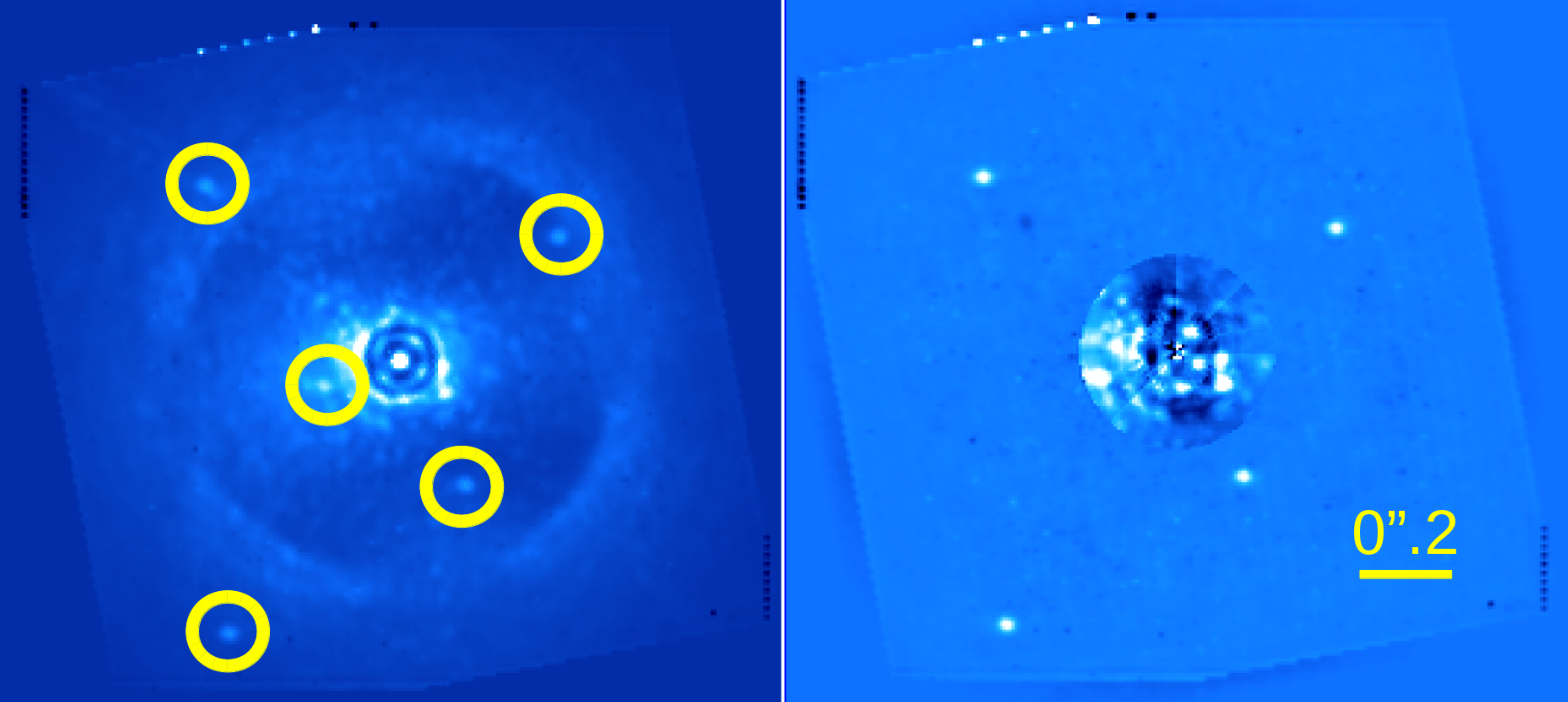}
\caption{\emph{Top:} Map of the synthetic T5-type planets (T$\sim$2000 K) of $1\times10^{-4}$ after injection in $H2$ filter (left) and SDI reduction (right) of IRDIS data. A slight misalignment of the Lyot stop causes the oblique stripe going through the image. \emph{Bottom:} Simulated T5-type (T$\sim$1100 K) planets at a contrast of $3\times10^{-5}$ injected in the IFS pre-reduced datacube (left) and after the SD reduction (right). In the central part of the image, the SD cannot work properly for the reasons explained in Sec.~\ref{sec:red}. The images show the 15th channel of IFS at $\lambda=1.09$~\mic.}
\label{f:map}
\end{center}
\end{figure}

After the injection, the raw IRDIS images were pre-reduced (background subtraction, flat-fielding, recentering, median recombination of the datacube, and rescaling) and the SDI was applied to minimize the speckles noise as described in Sect.~\ref{sec:red}. An image of the planets after injection and SDI reduction is shown in Fig.~\ref{f:map} (top). For the IFS, an example of the simulated planet inserted in the scientific datacube, after SD reduction, is also shown in Fig.~\ref{f:map} (bottom).

\section{Results}
\label{sec:results}

\subsection{Signal-to-noise ratio}
\label{sub:snr}

First, we determined in which cases the planets are detectable. We assumed that a planet was detectable if its signal-to-noise ratio (S/N) is above the value of 5\footnote{The values of 5, commonly used as a good false alarm rejector, will be effective in retrieving real detections with the addition of the ADI, while for this set of data we estimated that a level of 5 times the standard deviation of the signal still includes some false alarms, especially along the spiders projection. We empirically observed that the distribution of the S/N level of the pixels of our data is represented by an exponential function, while the expected distribution for an ADI processed image is expected to be Gaussian \citep{2008ApJ...673..647M}.}, with the S/N defined in this way:

\begin{itemize}
  \item for each pixel $p$ we calculate the integrated flux ($F_A$) in an aperture of radius 0.6$\lambda/D$ centered on $p$, and normalize it by the area of the aperture;
  \item we consider an annulus centered on $p$ with an inner radius of 3$\lambda/D$ and an outer radius of 5.5$\lambda/D$, to be close enough to the object and without including its flux;
  \item we consider another annulus centered on the star with inner and outer radii of $r-0.6\lambda/D$ and $r+0.6\lambda/D$, where $r$ is the distance between the star center and $p$;
  \item we calculate the background (median of the values) $bkg_B$ and the standard deviation $\sigma_B$ on two areas (B) that are the intersection of the two rings described above; and
  \item we define the S/N as: $S/N = (F_A-bkg_B)/\sigma_B.$
\end{itemize}

\begin{figure}[h]
\begin{center}
\includegraphics[width=0.3\textwidth]{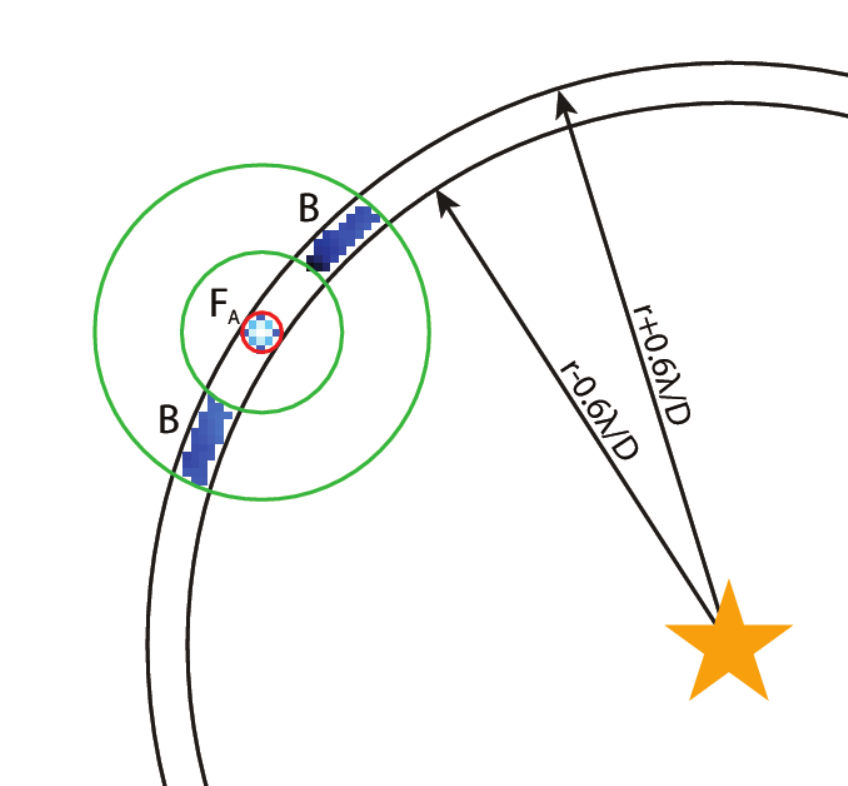}
\caption{Representation of the areas where the signal of the planet and the local noise are calculated. The zone $F_A$ (red circle) is the area where the signal is calculated, while the zone B (green circles) represents the area where the background and the standard deviation of the noise are calculated, as described in Sect.~\ref{sub:snr}.  }
\label{f:areas}
\end{center}
\end{figure}

\noindent A cartoon describing the different areas defined for the S/N calculation is shown in Fig.~\ref{f:areas}. The values of the radii and the choice of the zones is made to have a local value of the S/N, as the speckles noise is not homogeneous on the whole FoV and the structure of the spiders is visible.  

If we use the entire annulus centered on the star, the estimation of the SNR can change from some fraction of a percent to a maximum of 9\% for planets fainter than $10^{-4}$. We decided to use a local estimation of the background to avoid some instrumental features, such as the diffracted light from the spiders, which will have much less impact while exploiting the ADI technique. Once on the sky, especially for the cases close to the detection limits, we will explore different methods for the estimation of the SNR to reach the maximum rejection of the false alarms.
  
For IFS data, we had to distinguish the procedure used to find planets, where an aggressive method is used to create deep detection maps and the procedure to characterize them, where we know the position of the planet and we use optimized methods around the position of the object to recover its signal. This different approach between the detection part and the characterization of the detected object has to be kept in mind in general when dealing with direct imaging data \citep[see, e.g,][]{2014arXiv1409.6388P}. For the IFS detection maps, the aggressive method we used is the SD without any mask to protect the planet signal. The number of detected planets and the mean S/N is given for both instruments in Table~\ref{t:decpl} for each simulated separation and contrast. The optimized method to characterize the planets are the SD with a mask and KLIP, described in Sec.~\ref{sec:red}.

\begin{table*}
\caption{Number of detected planets out of 30, in function of the contrast (columns) and the separation (lines) for IRDIS and IFS subsystems. The mean S/N value for the detected planets is given in parenthesis. The definition of the S/N and how it is calculated for our data is described in Sec~\ref{sub:snr}.} 
\label{t:decpl}
\centering
\begin{tabular}{cllllllllll}
\hline
\hline
%\multicolumn{2}{ |c| }{ & 10$^{-3}$ &	3$\times$10$^{-4}$ &10$^{-4}$	&3$\times$10$^{-5}$& 10$^{-5}$} \\ 
 & \multicolumn{2}{ c }{10$^{-3}$}& \multicolumn{2}{ c }{3$\times$10$^{-4}$}& \multicolumn{2}{ c }{10$^{-4}$} &\multicolumn{2}{ c }{3$\times$10$^{-5}$} &\multicolumn{2}{ c }{10$^{-5}$}\\
%\backslashbox{Separation}{Contrast}  & 10$^{-3}$ &	$3\times10^{-4}$ &10$^{-4}$	&$3\times10^{-5}$& 10$^{-5}$\\
\hline
    & IRDIS &IFS & IRDIS &IFS & IRDIS &IFS & IRDIS &IFS & IRDIS &IFS \\

0\farcs20    &      30 (20)  &30 (70)	     &30 (9)	&30 (22)	&6 (7) &      27 (7)&0 (-)    &1 (6)  & 0 (-)  &  0 (-)\\
0\farcs35    &      30 (98)	&30 (347)	&30 (42)	&30 (129)	&30 (15)&      30 (46)&18 (6)   &30 (14) & 0 (-)  & 14 (8)\\
0\farcs50    &      30 (171)	 &30 (238)	&30 (64)	&30 (114)	&30 (22)&      30 (43)&23 (8)  &30 (12) & 5 (8)  & 5 (7)\\
0\farcs65    &      30 (260)	&30 (281)	&30 (93)	&30 (135)	&30 (31)&      30 (55)&29 (10)  &30 (17)& 6 (7) & 6 (8)\\
0\farcs80    &      30 (272)	&30 (285)	&30 (104)	&30 (130)	&30 (36)&      30 (50)&29 (12)  &26 (17)& 11 (7)& 11 (8) \\
\hline
\end{tabular}
\end{table*}

\subsection{Photometry with IRDIS}
\label{sub:phIRDIS}

Photometry on SDI data is difficult because, as shown in Fig.~\ref{f:map} (top, right), we have to deal with two peaks: one positive from the planet signal in $H2$ filter and one negative from the planet signal in $H3$ filter. The peaks can overlap and the {\it a priori} ratio between the fluxes in the two filters is unknown \citep[see, e.g,][]{2014arXiv1404.3502M}. Instead of attempting to measure the planet signal in the final SDI image, where degeneracies will necessarily occur because of the partial subtraction of the two peaks \citep{2014arXiv1404.3502M}, we have adopted a method based on the introduction of ``negative synthetic planets'' into the raw data \citep[see, e.g,][]{2011A&A...528L..15B}, adapted for SDI data, for which we expect that the degeneracies will be less limiting. This method is similar to what the routine \texttt{fitstars}, presented in \citet{1996AJ....112.1180T,2000AJ....119.2403T}, does to calculate the photometry of binaries. In the following paragraphs we describe this method in more detail.

We note $\mathcal{P}_{H2}$ and $\mathcal{P}_{H3}$ the PSFs of the planet in filters H2 and H3, respectively. Similarly, we note $\mathcal{M}_{H2}$ and $\mathcal{M}_{H3}$ the off-axis PSFs in filters H2 and H3, which we use to model the planet PSFs. Finally, we note $\mathcal{F}_{H2}$ and $\mathcal{F}_{H3}$ the numerical factors by which we need to multiply our models to obtain a representation of the planet PSFs. We write:

\begin{eqnarray*}
  \mathcal{P}_{H2} & = & \mathcal{F}_{H2}.\mathcal{M}_{H2}, and \\
  \mathcal{P}_{H3} & = & \mathcal{F}_{H3}.\mathcal{M}_{H3}.
\end{eqnarray*}

\noindent Additionally, we assume that the planet PSFs are located at an unknown position ($\Delta$x,$\Delta$y) with respect to the star position. This value is independent from the filter.

The photometry and astrometry method for IRDIS attempts to determine the values of $\mathcal{F}_{H2}$, $\mathcal{F}_{H3}$, $\Delta$x and $\Delta$y by subtracting the planet signal from the raw data. For the practical implementation, we used the Levenberg-Marquardt least-squares fitting routine \texttt{MPFIT} \citep{2009ASPC..411..251M} with a custom function that takes as an argument the current values of the parameters being fitted: ($f_{H2}$,$f_{H3}$) for the fluxes, and ($\delta$x,$\delta$y) for the position. The function performs the following steps:

\begin{enumerate}
  \item creation of a model of the planet from the model of the planet, scaled with the current flux values:
    \begin{eqnarray*}
      m_{H2} & = & f_{H2}.\mathcal{M}_{H2} \\
      m_{H3} & =  & f_{H3}.\mathcal{M}_{H3}.
    \end{eqnarray*}
  \item subtraction of $m_{H2}$ and $m_{H3}$ in the raw H2 and H3 data, respectively, shifting them around the true position $\Delta$x,$\Delta$y at each iteration of the fit;
  \item application of the SDI procedure described in Sect.~\ref{sec:red} to obtain a final SDI image $\mathcal{I}_{SDI}$;
  \item return the residual variance in a zone ($\mathcal{C}$) made of the union of two circular apertures of diameter $1.5\lambda/D$ centered at the location of the planet in H2, and in H3 after the spatial rescaling. This residual variance is used for the least-squares minimization. 
\end{enumerate}

\noindent When a local or global $\chi^2$ minimum is found, the current values of $f_{H2}$, $f_{H3}$, $\delta$x, and $\delta$y are assumed to represent good estimations of $\mathcal{F}_{H2}$, $\mathcal{F}_{H3}$, $\Delta$x, and $\Delta$y respectively, i.e., for the photometry we write:

\begin{eqnarray*}
  \mathcal{P}_{H2} & \simeq & f_{H2}.\mathcal{M}_{H2} \\
  \mathcal{P}_{H3} & \simeq & f_{H3}.\mathcal{M}_{H3}.
\end{eqnarray*}
%\newpage
As mention above, this method could be affected by degeneracies for planets very close to the central star, where the peaks at both wavelengths overlap significantly. Future work will be devoted to improving this technique and precisely identifing its limitations, but the results detailed below show that it is promising even for planets as close as 0\farcs2.

Photometry measurements were performed for all the synthetic planets introduced in the data, as detailed in Sect.~\ref{sec:fp}. For the starting point of $f_{H2}$ and $f_{H3}$, we assumed a value 10\% above the true values $\mathcal{F}_{H2}$ and $\mathcal{F}_{H3}$. This choice is arbitrary, but we performed tests to verify that varying this starting point did not have significant impact on the result. We also tested the impact of varying the size of zone $\mathcal{C}$ for the minimization of the residual variance, and verified that there is no impact for the size between 1 and 2$\lambda$/D.

\begin{figure}
\begin{center}
\includegraphics[width=0.24\textwidth]{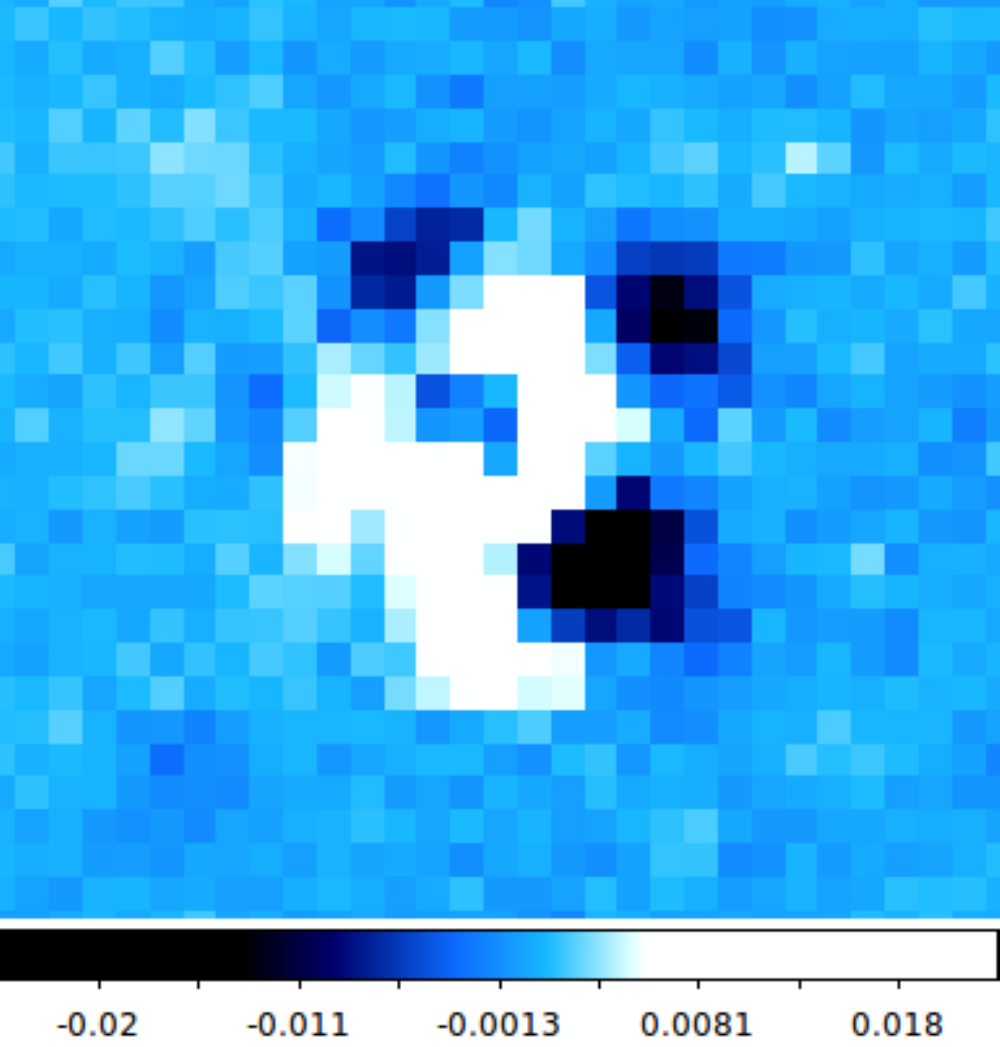}
\includegraphics[width=0.24\textwidth]{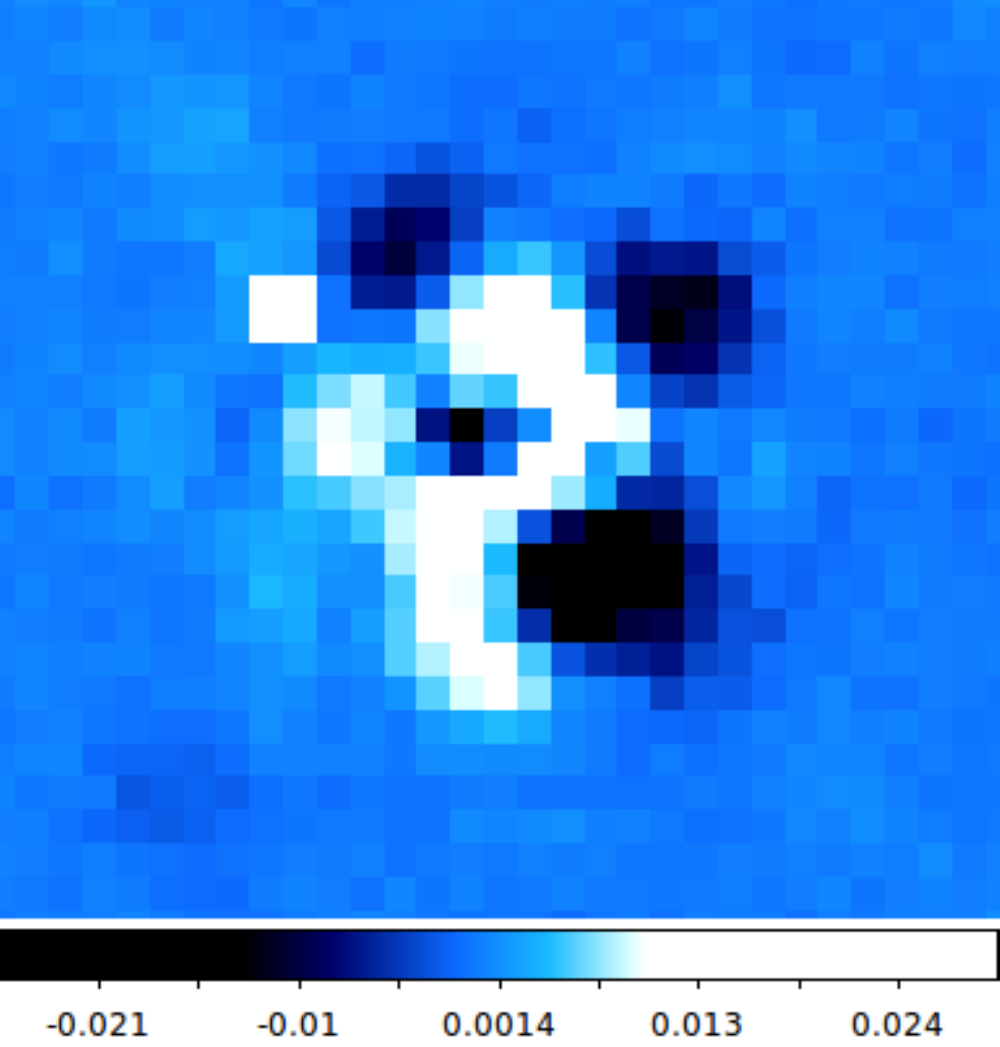}
\caption{Residual from the subtraction of the off-axis PSF taken right after the sequence and a different PSF taken some months before for H2 filter (right) and H3 (left). The peak-to-peak variation is about the 5-6\% of the PSF flux and the standard deviation of the central zone (12$\times$12 px) is 1\%.}
\label{f:res}
\end{center}
\end{figure}

To have a realistic prediction of what we could measure when conditions would vary significantly, we performed this method using two different off-axis PSFs as a model:
\begin{itemize}
\item Ideal case: the PSF of the planets can be well represented by off-axis PSFs $\mathcal{M}_{H2}$ and $\mathcal{M}_{H3}$ acquired right after the coronagraphic sequence, which means that the system and the atmospheric conditions are stable during the sequence.
\item Variable PSF case: the model used to represent the PSF of the planets is an off-axis PSF taken some months before the scientific exposure, which means that the system and the atmospheric conditions changed during the exposure.
\end{itemize}
The residual between the two PSFs are shown in Fig~\ref{f:res}. The peak-to-peak variation of the difference between the two PSFs is about the 5-6\% of the PSF flux and the standard deviation of the central zone (12$\times$12 px) is 1\%.
A realistic case on the sky should lie in between the two boundaries, depending on the stability of the conditions during the scientific acquisition. Future on sky tests will be dedicated to the study of the variability of the PSF shape during the scientific exposures.  

\begin{figure}
\begin{center}
\includegraphics[width=0.48\textwidth]{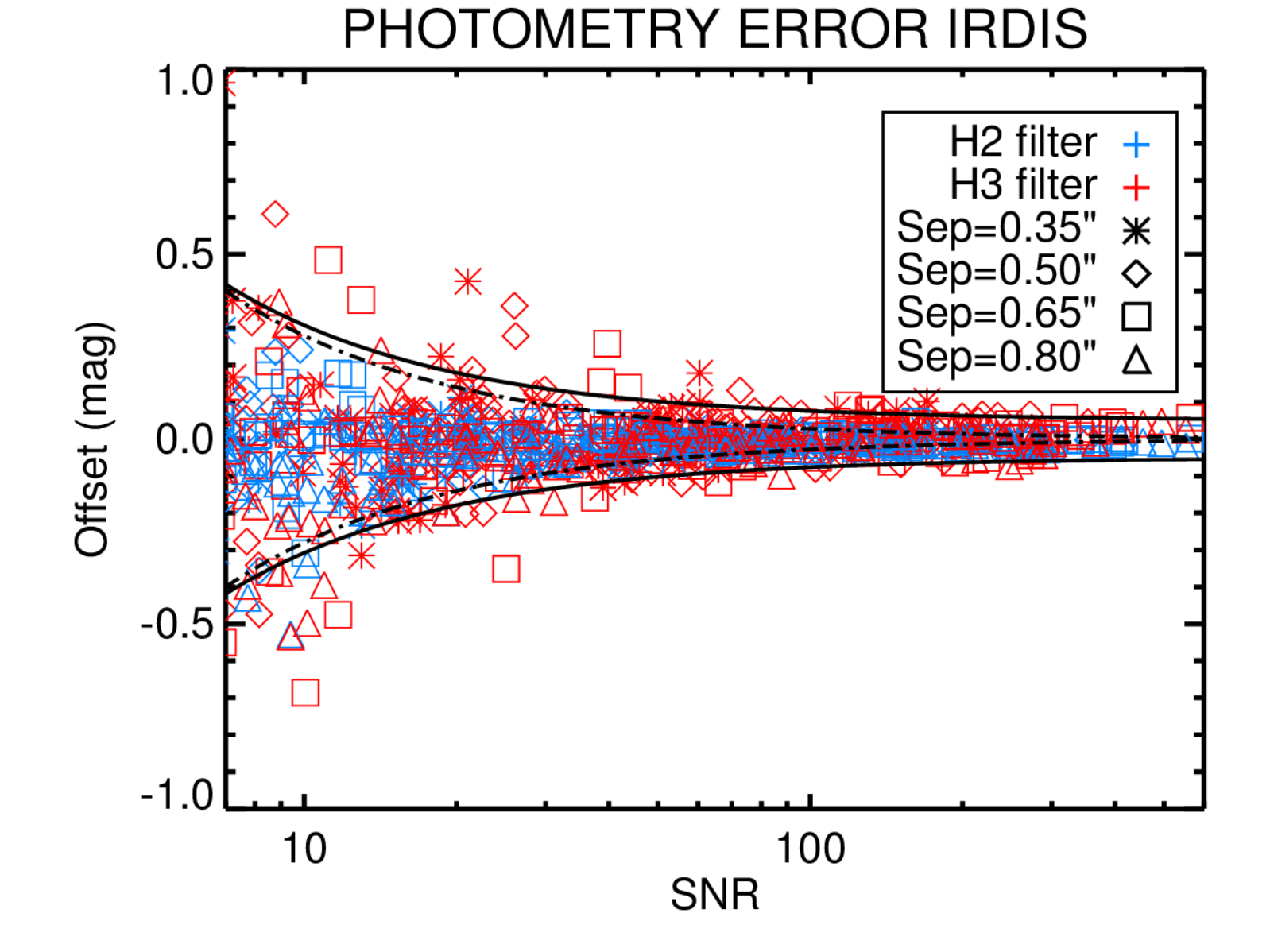}
\caption{Plot of the difference between nominal and measured values of the magnitude versus the S/N of each planet with contrasts from 10$^{-5}$ to 10$^{-3}$ and different separations from the host star for the SDI image. The dashed black line represents Eq.~\ref{eq:photoirdis2}. Different separations from the host star are represented with different symbols.}
\label{f:photo_irdis}
\end{center}
\end{figure}

The offsets in magnitude between the nominal and measured flux versus the S/N of the planets are shown in Fig.~\ref{f:photo_irdis} for the extreme case. The trend of the error for the different PSF case in function of the S/N is calculated as: 

\begin{center}
\begin{equation}
    \sigma = 0.07 + \frac{2.6}{S/N} mag
\label{eq:photoirdis2}
\end{equation}
\end{center}
\noindent and is plotted as a solid line in Fig.~\ref{f:photo_irdis}, while for the ideal case is calculated as:

\begin{center}
\begin{equation}
    \sigma =  \frac{2.8}{S/N} mag
\label{eq:photoirdis}
\end{equation}
\end{center}  

\noindent and overplotted as a dashed line.

This can demonstrate that for the IRDIS photometric measurement, the variability of the PSF during the exposure does not have a big impact, as the photometric fit considers residuals inside an aperture.

\begin{figure*}
\centering
\includegraphics[width=0.46\textwidth]{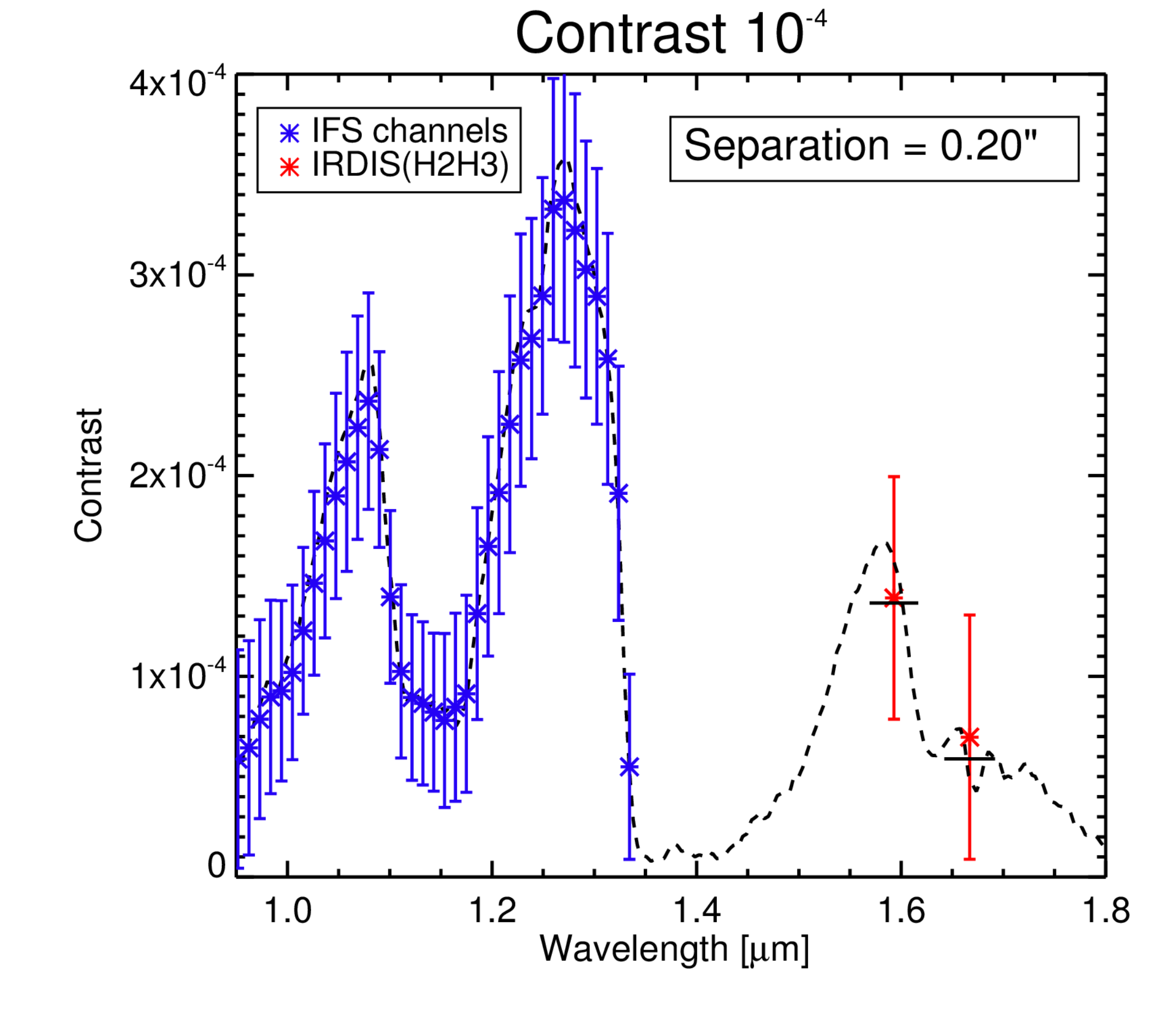}       %%%%%%cambiare figura
\includegraphics[width=0.46\textwidth]{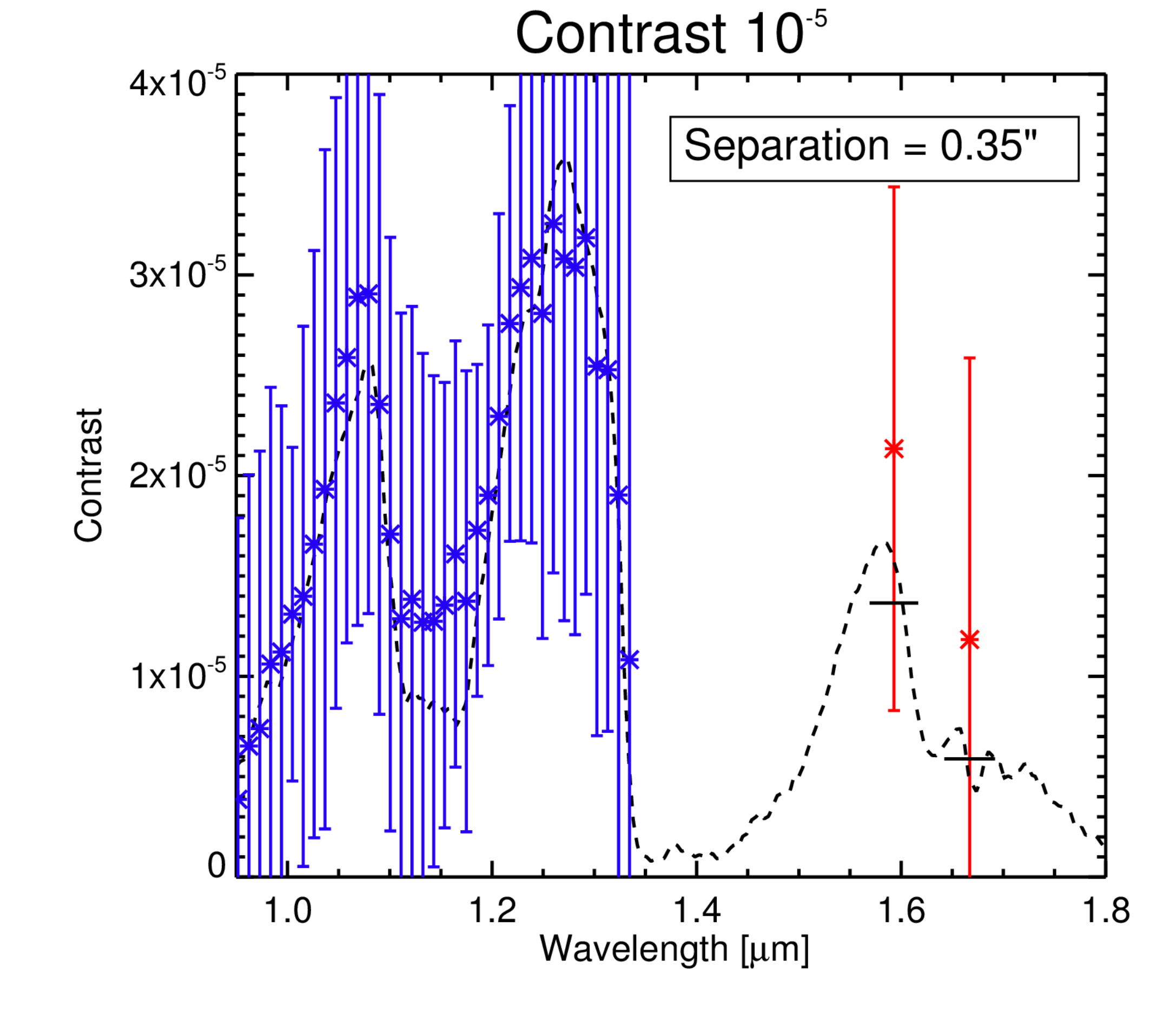}
\includegraphics[width=0.455\textwidth]{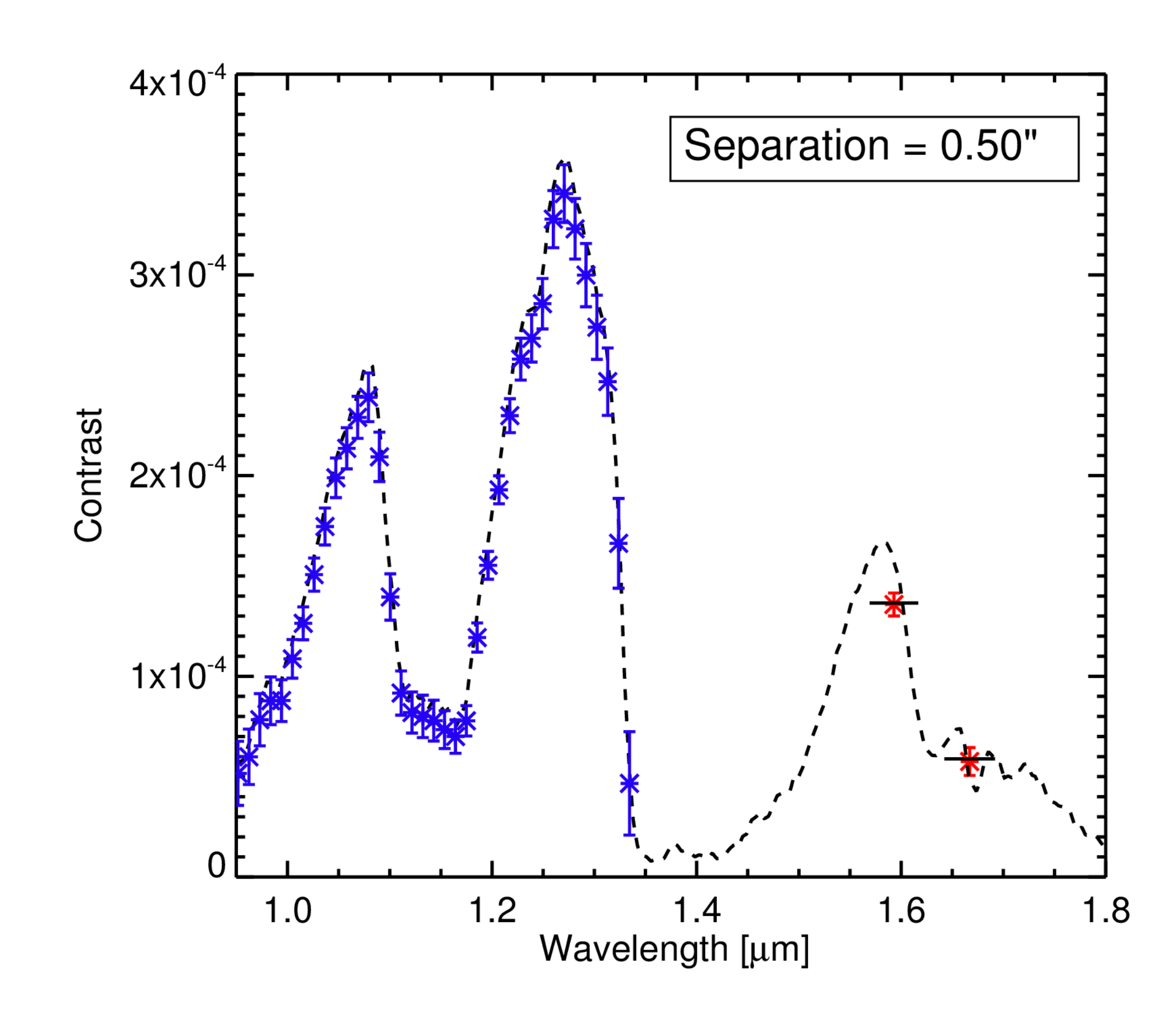}
\includegraphics[width=0.455\textwidth]{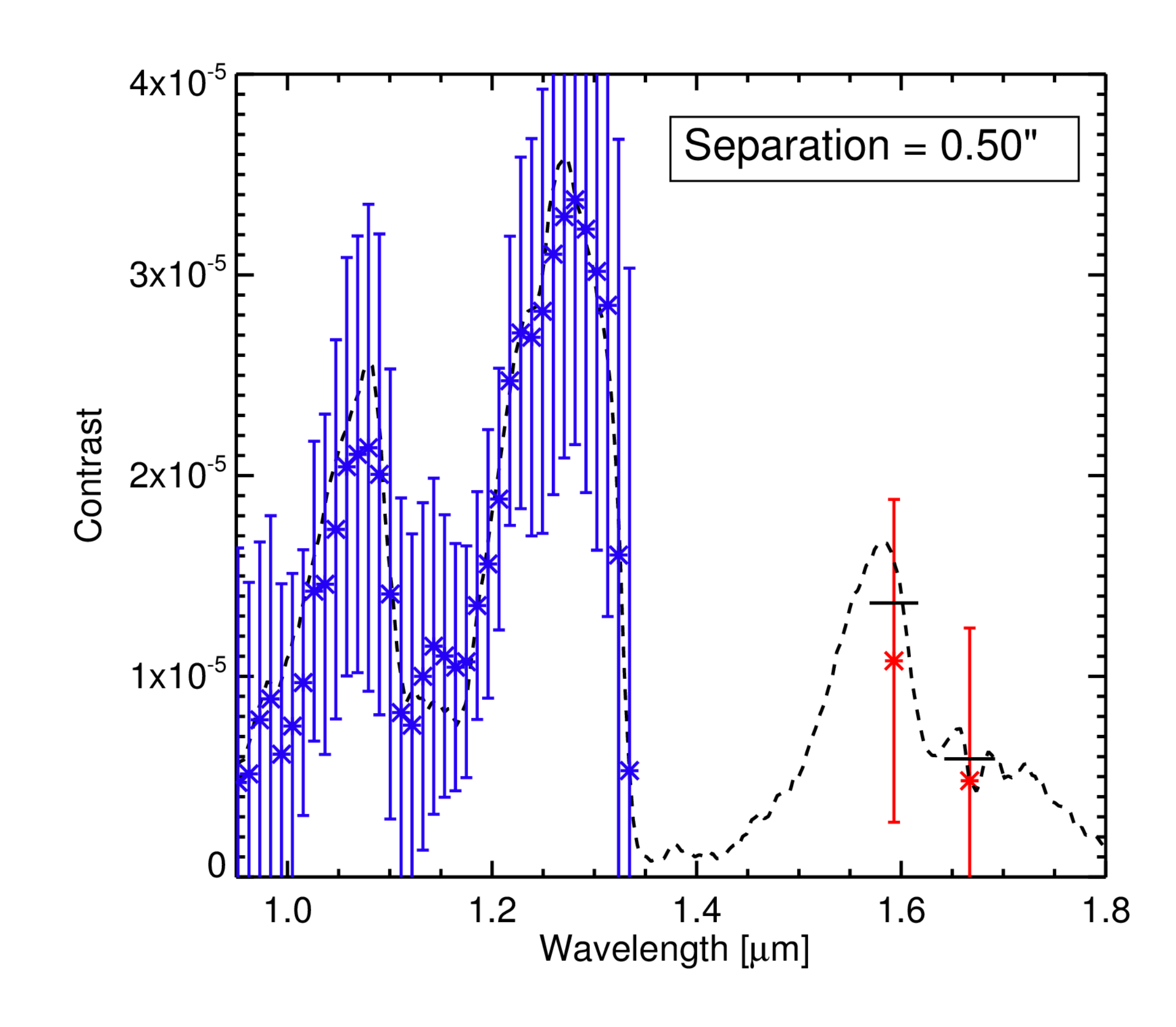}
\includegraphics[width=0.455\textwidth]{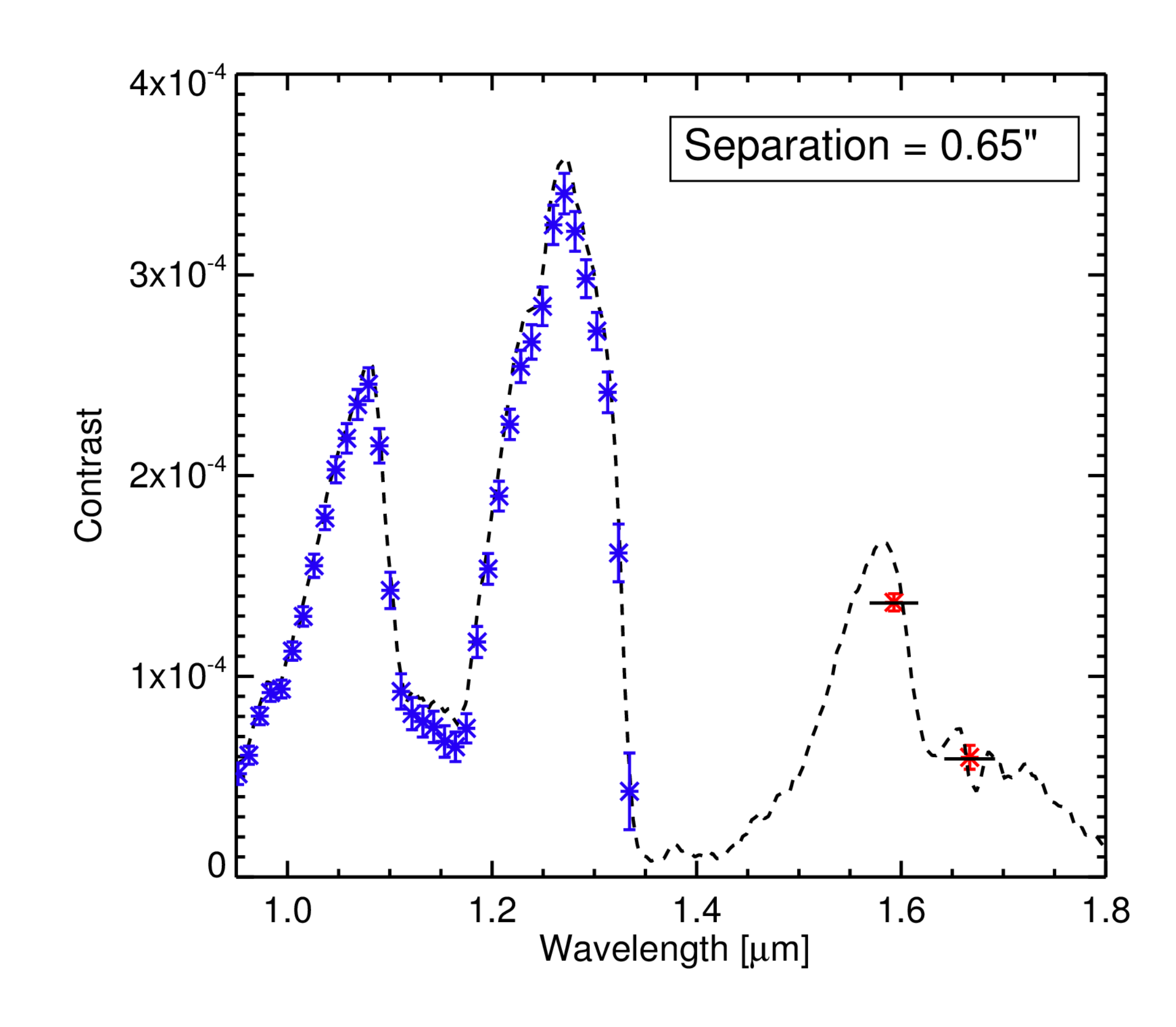}
\includegraphics[width=0.455\textwidth]{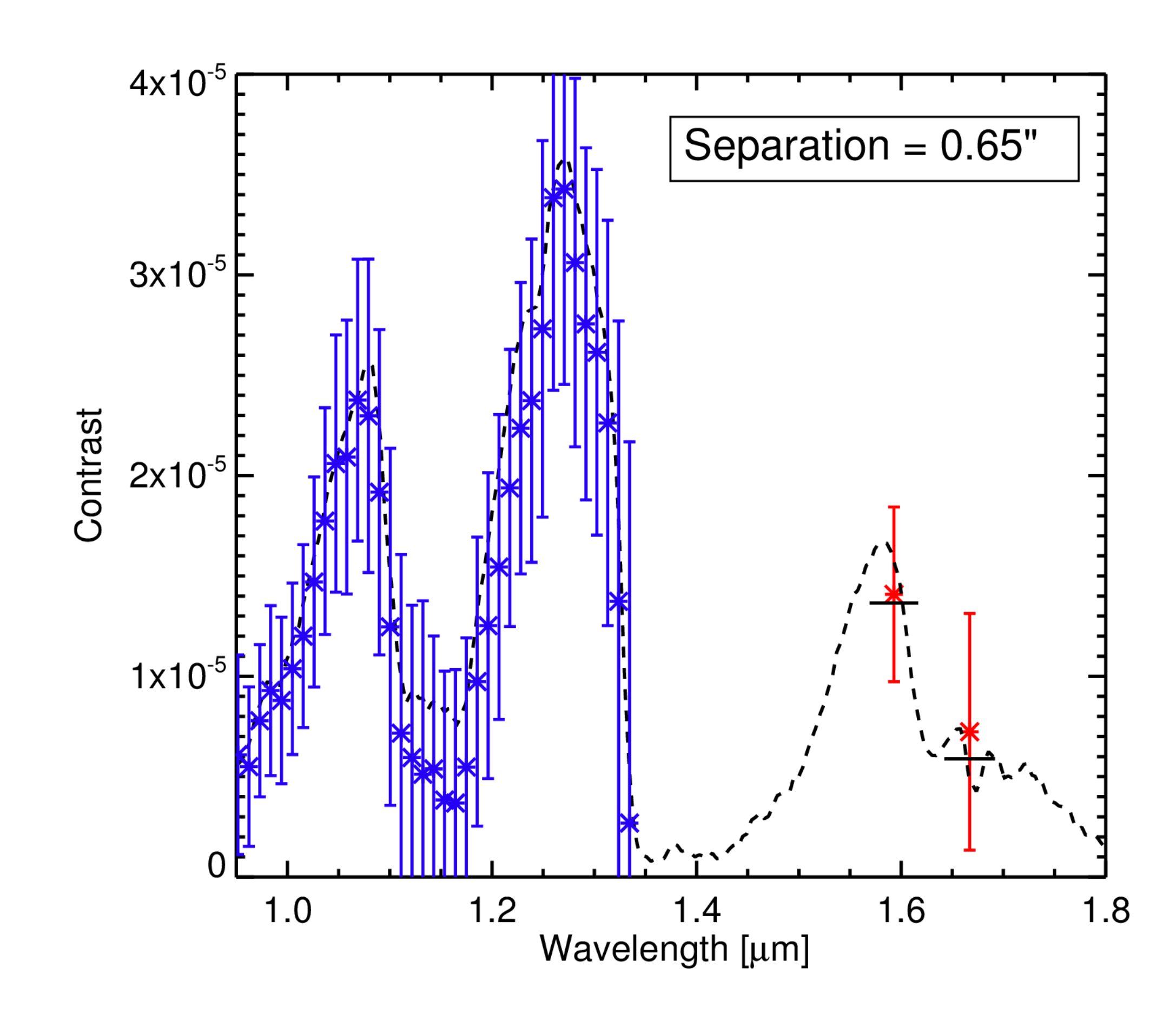}
\caption{Spectral extraction of a T5 (T$\sim$1100 K) model spectrum (black line) for planets at different separations (from top to bottom) and contrast from the star ($10^{-4}$ on the left, $10^{-5}$ on the right). The blue points represent the IFS photometry for each channel (reduced with the SD technique), while the red ones represent the flux measurement in the two IRDIS filters $H2$ and $H3$. For IRDIS data, the black horizontal line represents the therotical value of the photometry; each line covers the bandpass of the filters.}
\label{f:photo}
\end{figure*}

Examples of the two IRDIS photometry channels measured with the different PSF as a model for a T5-type spectrum, for two different contrasts and three separations from the host star are shown in red in Fig.~\ref{f:photo}. The theoretical value of the photometry for each filter is also represented as a black horizontal line. The error bars are calculated as the standard deviation of the measurements of each planet in 30 different positions. 

\subsection{Astrometry with IRDIS}
\label{sub:astroIRDIS}

\begin{figure}
\begin{center}
\includegraphics[width=0.48\textwidth]{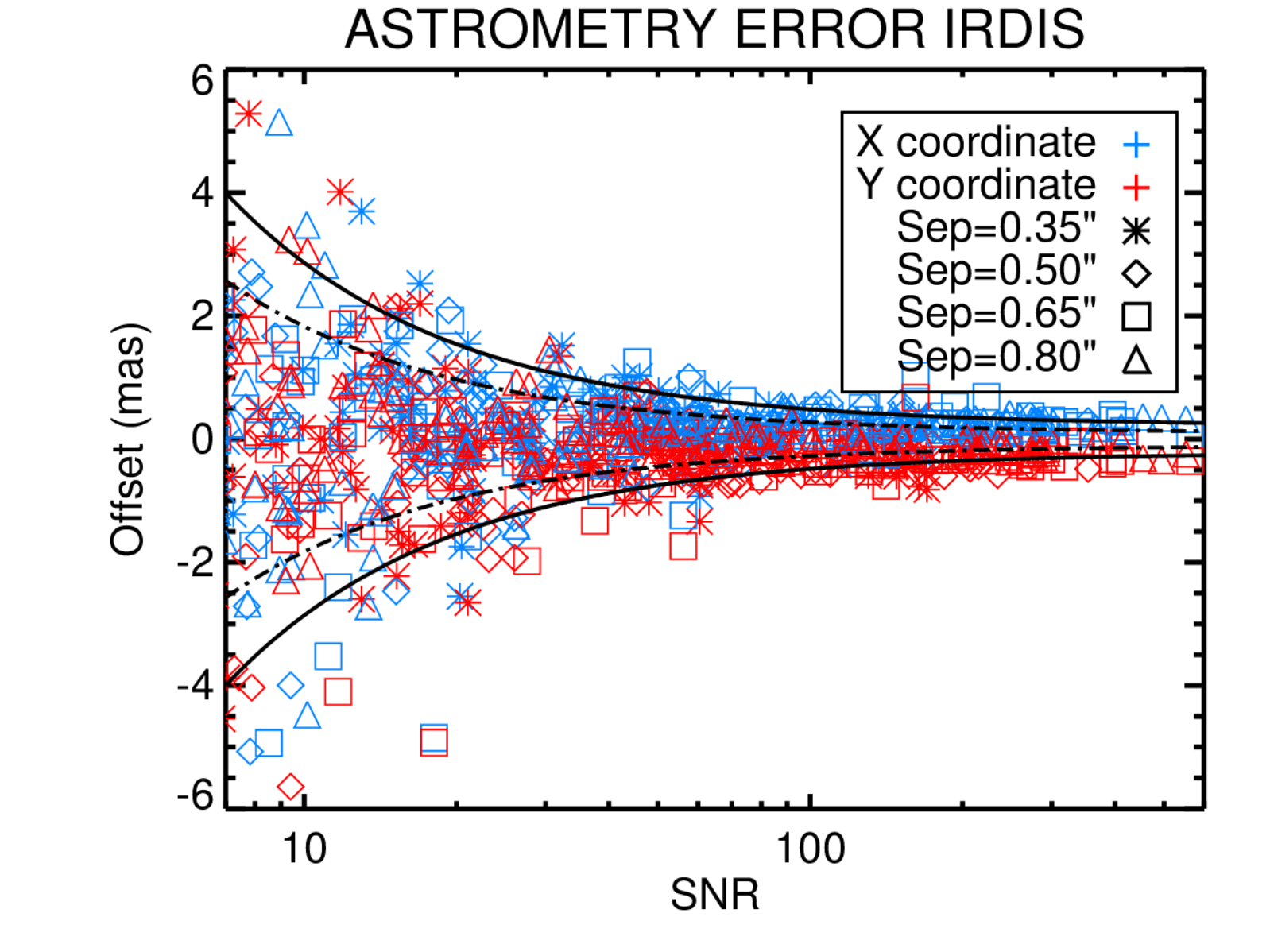}
\caption{Plot of the difference between nominal and measured astrometric values for the two coordinates versus the S/N of each planet with contrasts from 10$^{-5}$ to 10$^{-3}$ and separations from 0\farcs35 to 0\farcs80 in the SDI image. The dashed black line represents Eq.~\ref{eq:astroirdis}. Different separations from the host star are represented with different symbols.}
\label{f:astroerr01}
\end{center}
\end{figure}

 For the astrometric analysis, we used the same least-squares method presented in Sect.~\ref{sub:phIRDIS} to calculate the position of the simulated planets. The fit is done simultaneously with the photometric measurement, and the PSFs are shifted from their original position by 8 mas. The results for the case where the model PSF is different than the off-axis PSFs are presented in Fig.~\ref{f:astroerr01}, where we plot the astrometric errors on both axes as a function of the S/N. In this figure, only detected planets (481 out of 600) are shown. Moreover, the planets at the closest separation of 0\farcs2 are not taken into consideration for this analysis because of the large uncertainties obtained for them.

To estimate the final error in the astrometry we calculated the standard deviation $\sigma$ of these values and extrapolated their trend. The expected astrometric error for the relative planet-star position is shown in function of the S/N of the planets in Fig.~\ref{f:astroerr01}. The curve overplotted (solid black line) is given by:  

\begin{center}
\begin{equation}
    \sigma =  0.2 + \frac{26.4}{S/N}~mas,     %     0.218530      26.4462
\label{eq:astroirdis}
\end{equation}
\end{center}  

\noindent for the different PSF case, while for the ideal case is given by:

\begin{center}
\begin{equation}
    \sigma =  0.1 + \frac{17.3}{S/N}~mas,     %     0.218530      26.4462
\label{eq:astroirdis2}
\end{equation}
\end{center} 

\noindent and plotted as a dashed line in Fig.~\ref{f:astroerr01}. In this case, the minimization of the residuals is biased by the different shapes of the PSF and there is a rigid shift in the coordinates of the centroid found by the fit. A realistic case should fall in between the two boundaries described by Eq.~\ref{eq:astroirdis} and \ref{eq:astroirdis2}.

From these results, we predict a relative astrometric precision for IRDIS from a few tenths to $\sim$3 mas depending on the S/N of the candidate. This error also includes the error in the recentering introduced during the measurement of the waffle images and the shift of each frame, but it does not include the errors related to the absolute position of the planet (true north determination, distorsion, platescale orientation, ...). 

\subsection{Photometry with IFS}
\label{sub:phIFS}

To estimate the errors on the photometry measurements with IFS after SD (as presented in Mesa et al., submitted) and KLIP reductions we performed PSF-fitting photometry on the data where we injected planets. We made this choice because there are 39 spectral channels for IFS, and a method similar to that used for IRDIS analysis would not easily converge. We first reduced the dataset presented in Sec.~\ref{sec:fp} with SD and KLIP method separately and then we applied the PSF-fitting on the planets. The position of the planets is found by \texttt{MPFIT}, as described in Sec~\ref{sub:astroIFS}.

The procedure consisted in comparing a different off-axis PSF for each spectral channel with the PSFs of the planets. The contrast is calculated pixel by pixel after the alignment of the centers of the PSFs of the star and the planet. The final result is calculated as the weighted median of all the pixels inside a radius of 1.2$\lambda/D$, as described in the following equation:

\begin{center}
\begin{equation}
    Contrast = \iint{ \frac{ \frac{flux(x,y)}{model(x,y)}\left(\frac{model(x,y)}{k}\right)^2} { noise_{bkg} +\frac{model(x,y)}{k} }dxdy}.
\label{eq:cont_formula}
\end{equation}
\end{center}

\noindent Here $k$ is a normalization constant, $noise_{bkg} = (bkg_B)^2/2$ (see Sect.~\ref{sub:snr}), and $flux(x,y)$ is the value converted in ADU/s of the pixel (x,y) after subtraction of the background estimated in the same area $B$ as described in Sect.~\ref{sub:snr} (Fig.~\ref{f:areas}). In this manner, the central part of the PSF has a stronger impact than the wings. The same procedure was performed using a 2-d Moffat function that reconstructs the different off-axis PSF of the star, obtaining the same results. 
 
As the procedures of the SD and the PCA generally cause a loss of flux of the planets, we tried to reduce this problem using a mask that protects the zones where the planets sit during the reduction. To calculate the errors on photometry for each channel of IFS, we calculated the standard deviation of the residuals of all the planets with the same separation, flux, and wavelength.

\begin{figure*}
\centering
\includegraphics[width=0.4\textwidth]{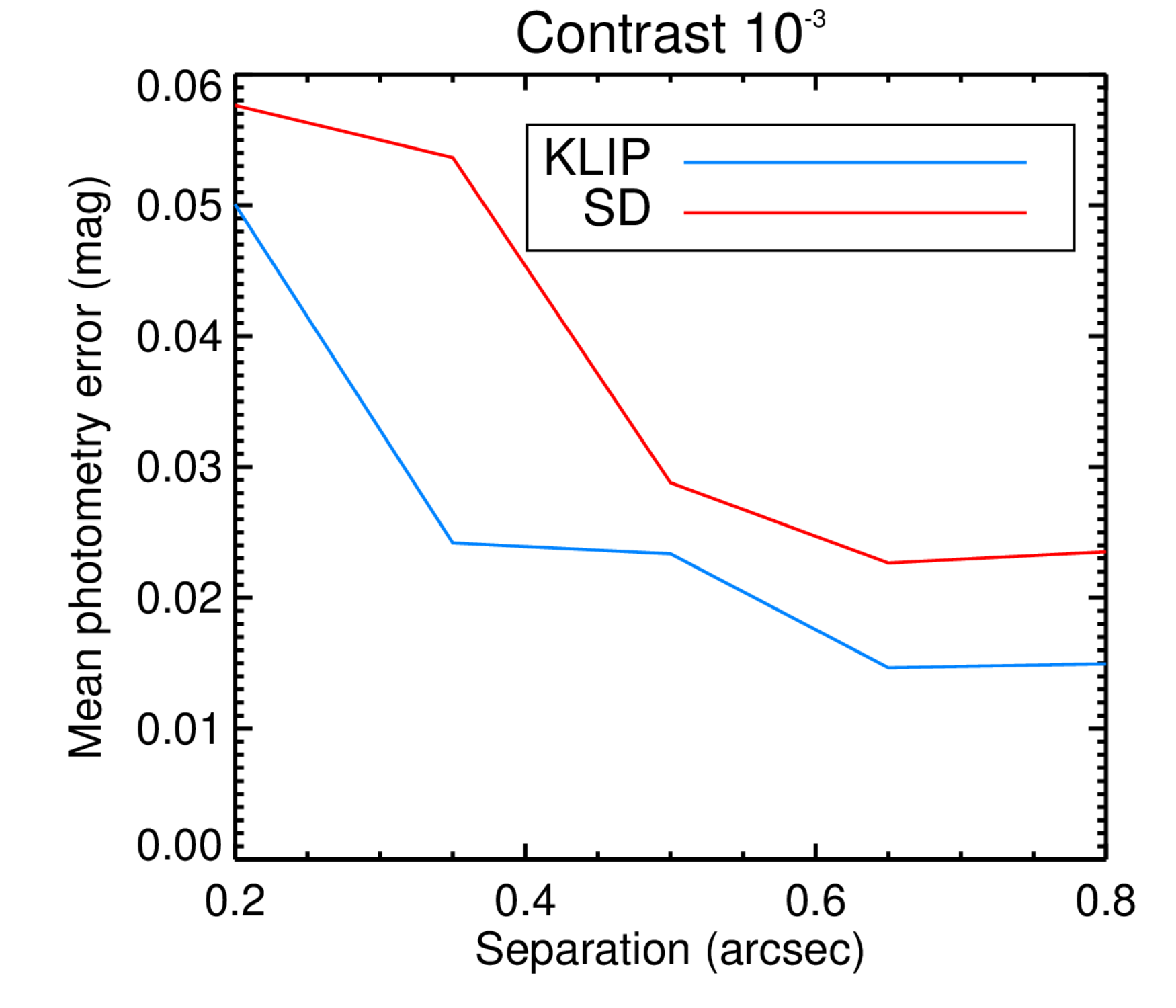}
\includegraphics[width=0.4\textwidth]{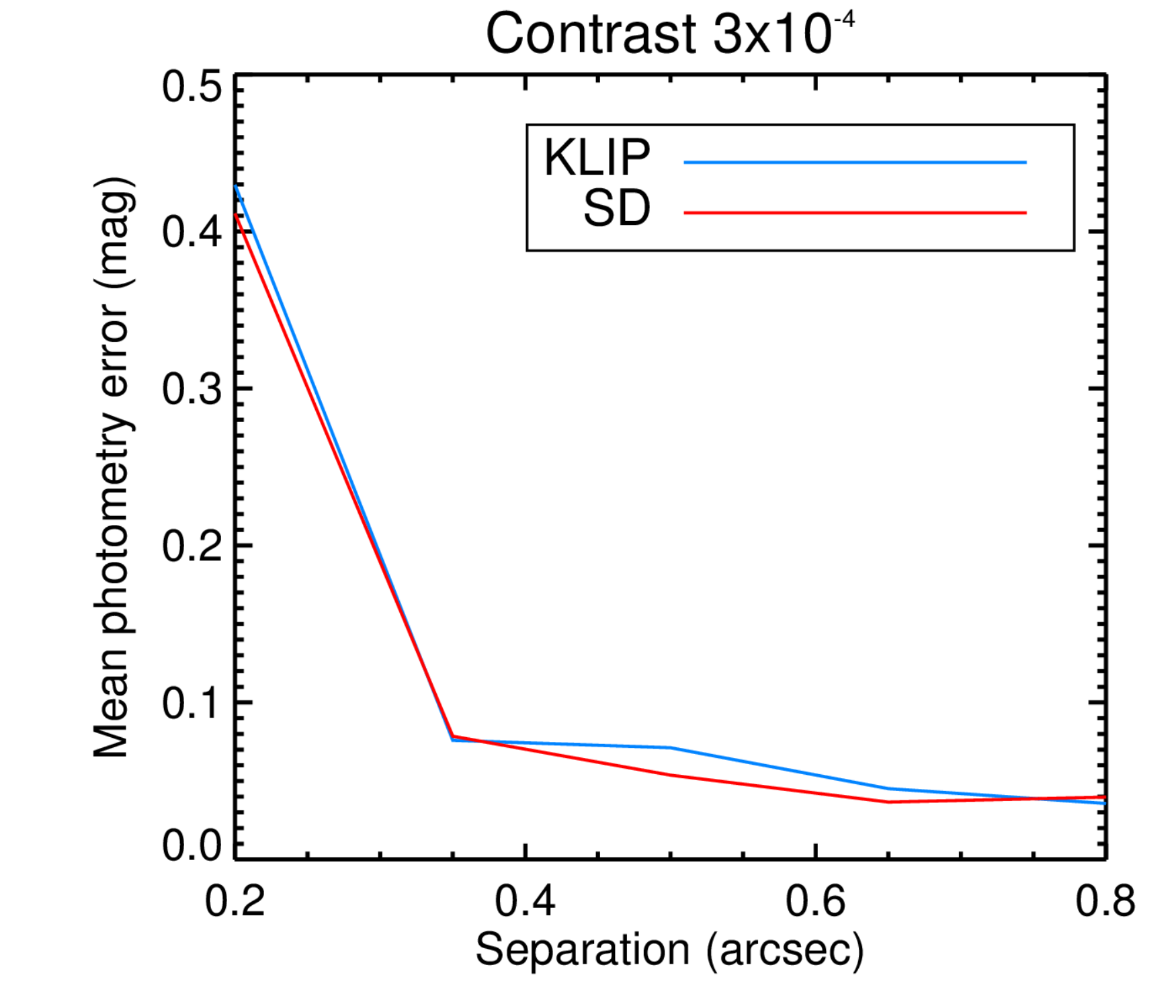}
\includegraphics[width=0.4\textwidth]{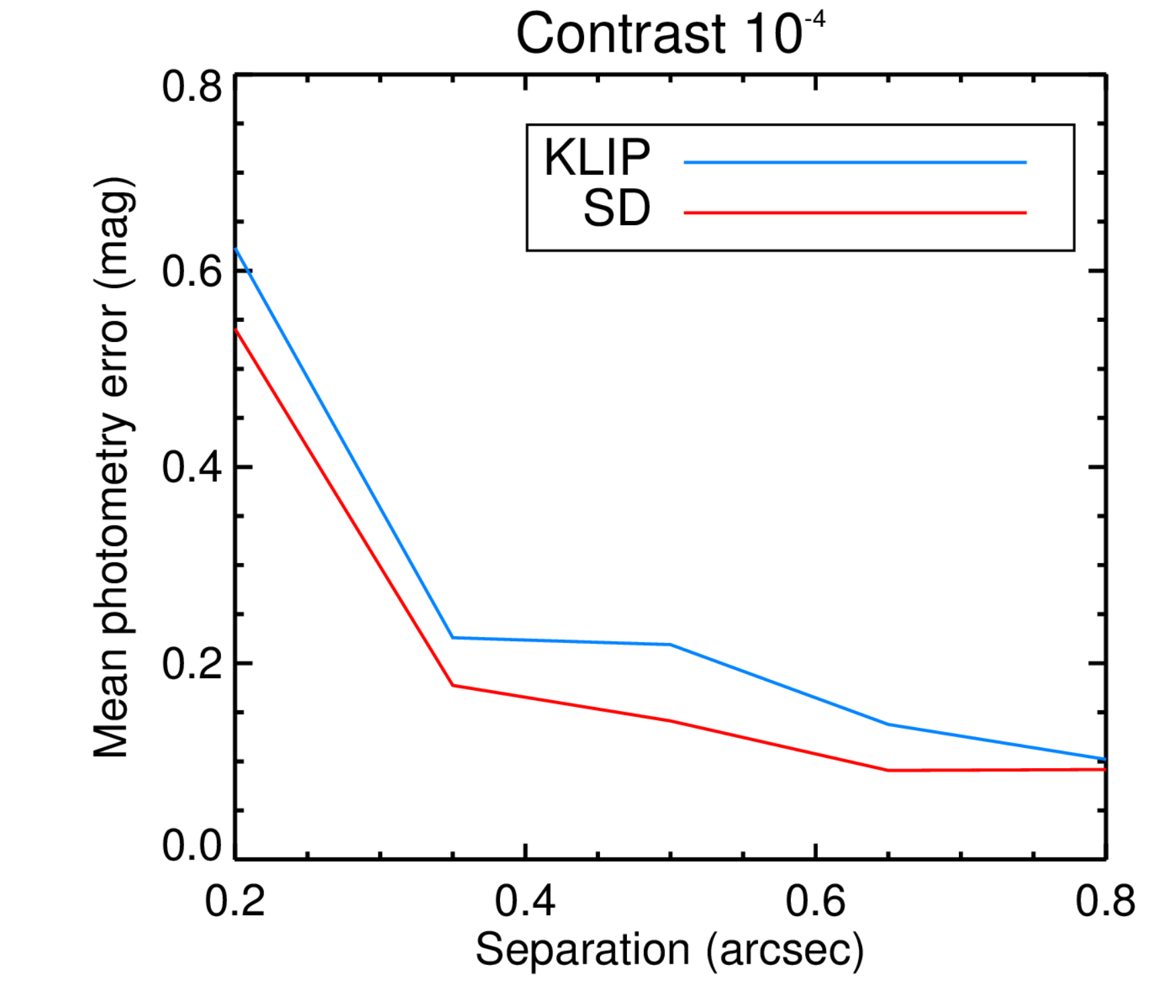}
\includegraphics[width=0.4\textwidth]{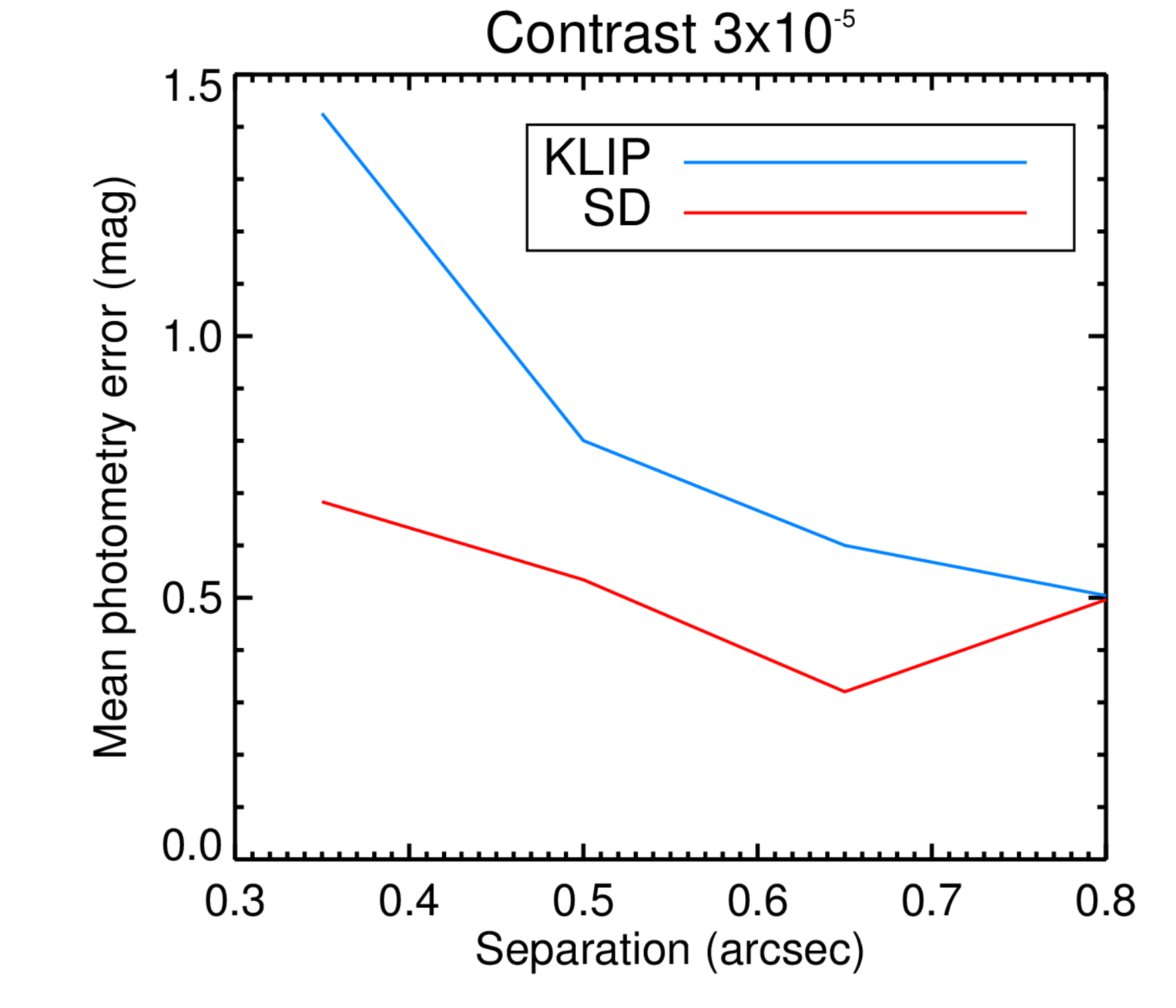}
\caption{Plots of the expected photometric errors for different contrasts as a function of separation from the central star. Results obtained with KLIP (red line) and SD (blue line) are compared. Each point represents the median of the errorbar on the 39 IFS channels. Only detected planets are considered in this analysis. Very few planets with contrast of 10$^{-5}$ are detected, as shown in Table~\ref{t:decpl}, so the corresponding plot is not shown.}
\label{f:k_sd}
\end{figure*}

In order to improve KLIP results we used an implementation of the forward modeling presented in \citet{2012ApJ...755L..28S}. To evaluate which method is better to use as a function of separation and contrast, we calculated the standard deviation of the photometry offsets obtained reducing the data with the SD and KLIP techniques. The results of this analysis are presented in Fig.~\ref{f:k_sd}. From this evaluation, we determined that in general KLIP works better for brighter objects and that its results are comparable with the results of the SD. Nevertheless, we expect that KLIP will greatly improve with the addition of the FoV rotation, which permits us to expand the part of the library without the signal from the object itself.

Examples of the extracted spectra for T5-type planets at contrasts of $10^{-4}$ and 10$^{-5}$ and separations within the range of 0.2--0.8 arcsec are shown in Fig.~\ref{f:photo}, with error bars reflecting the dispersion of the results obtained with synthetic planets at different positions within the image. 

As for IRDIS case, the use of a PSF taken in different conditions does not impact the spectrum extraction.

\subsection{Astrometry with IFS}
\label{sub:astroIFS}

For astrometric measurements, we used the same procedures described as for IRDIS (see Sect.~\ref{sub:astroIRDIS}) with the only difference that the final step of the analysis, the minimization of the standard deviation in the aperture, is done on the median of all the IFS spectral channels.

After using the off-axis PSF taken right after the sequence and a PSF taken with different observation conditions as a model, as we did for IRDIS, we found that for the variable PSF case the method of the ``negative synthetic planets'' introduced larger error than finding the centroid of the planets with \texttt{MPFIT2DPEAK}, a routine that fit the image with a 2-D Gaussian profile. The difference of the two PSFs, for the case of IFS, has a peak-to-peak variation of 31\% and a standard deviation of the 3\%. Also, the wings of the PSFs are asymmetric and this fact could strongly affect the astrometry measurements. This is probably induced by the dependence of the IFS wavelength calibrations with time, as the two PSFs have been taken well separated in time (some months). As the IFS instrument is so sensitive to the calibrations we expect that when observing on the sky, where calibrations are taken during the same night of the observation sequence, the shape of the PSF should be more stable. 

We performed astrometric measurements both on SD and KLIP reduced datacubes. The expected error is calculated taking into account the dependency of the standard deviation of the offsets on the S/N of the candidates. The total numbers of detected planets (S/N $>$ 5) are 523 and 441 out of 600 for the SD (with mask) and KLIP reductions, respectively. Planets with separation of 0\farcs20 are not considered in this analysis because of the very low number of detected planets at this separation. 

\begin{figure}[h]
\begin{center}
\includegraphics[width=0.48\textwidth]{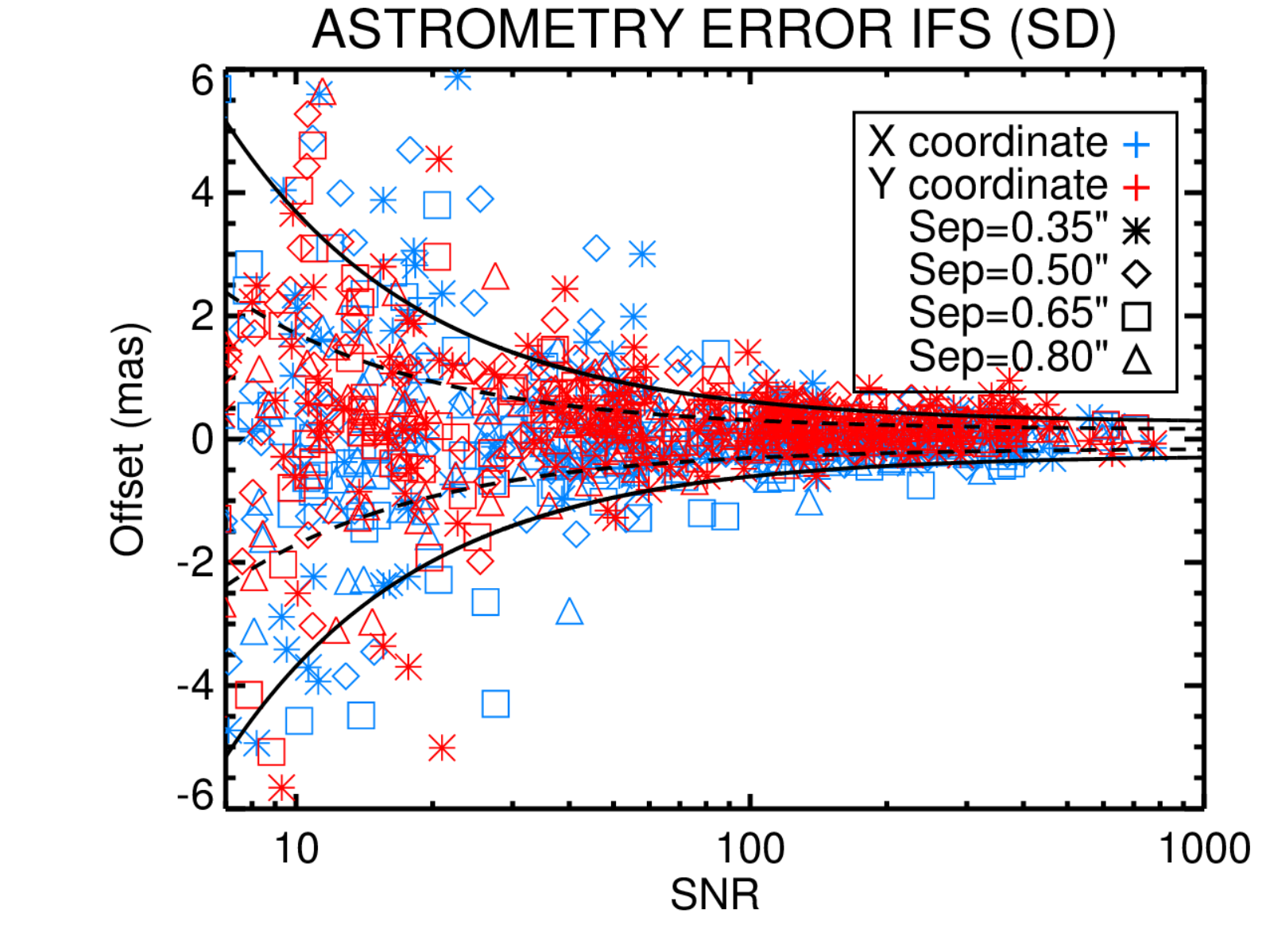}
\caption{Plot of the difference between nominal and measured astrometric values for the two coordinates versus the S/N of each planet with contrasts from 10$^{-5}$ to 10$^{-3}$ and separations from 0\farcs35 to 0\farcs80 in the IFS datacube after SD reduction. The dashed black line represents Eq.~\ref{eq:astroifs_sd}. Different separations from the host star are represented with different symbols.}
\label{f:astroerr_snr_ifs_sd}
\end{center}
\end{figure}

\begin{figure}[h]
\begin{center}
\includegraphics[width=0.48\textwidth]{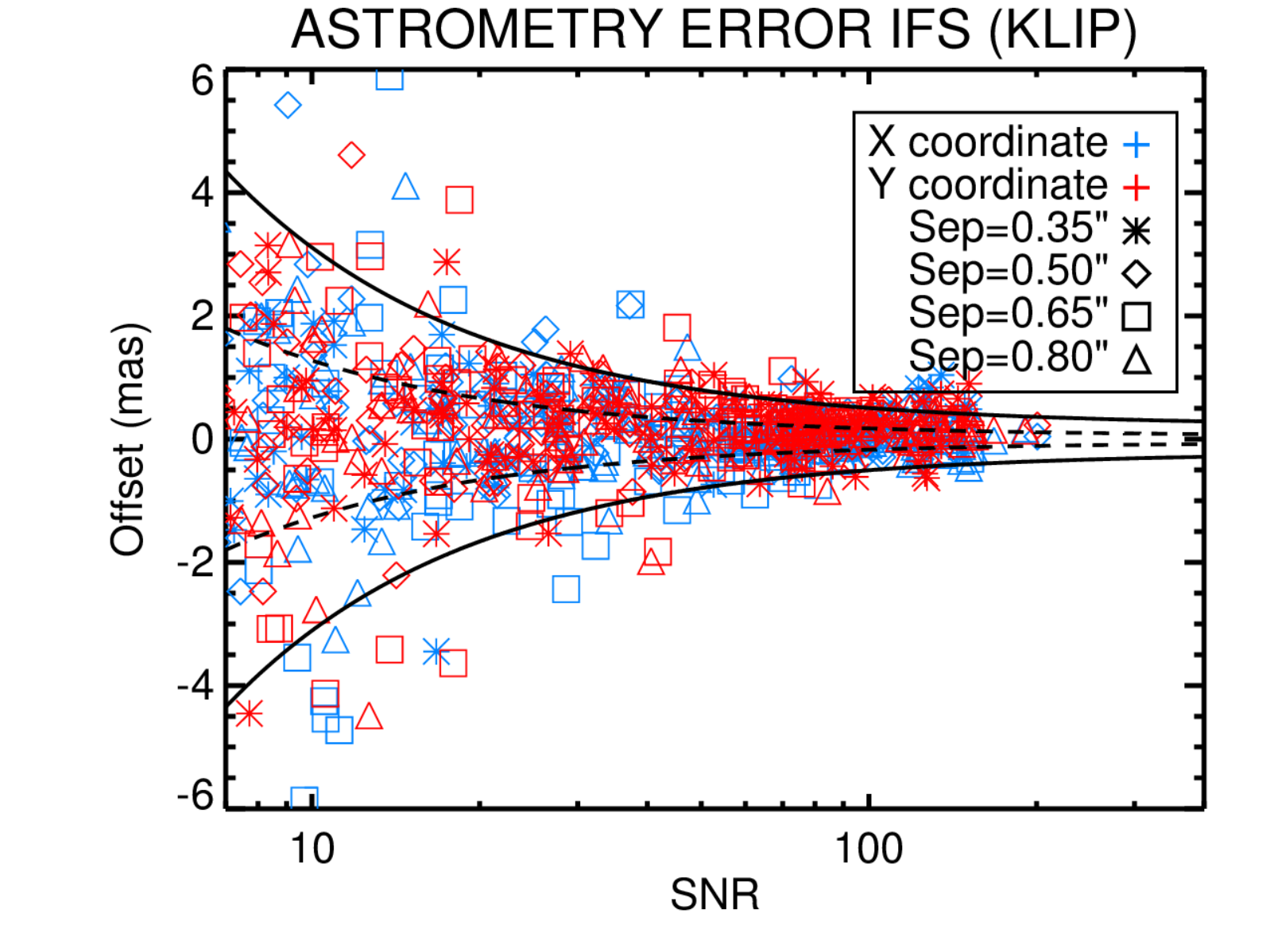}
\caption{Plot of the difference between nominal and measured astrometric values for the two coordinates versus the S/N of each planet with contrasts from 10$^{-5}$ to 10$^{-3}$ and separations from 0\farcs35 to 0\farcs80 in the IFS datacube after KLIP reduction. The dashed black line represents Eq.~\ref{eq:astroifs}. Different separations from the host star are represented with different symbols.}
\label{f:astroerr_snr_ifs_klip}
\end{center}
\end{figure}

We represent the astrometric offsets along the two cartesian coordinates for the SD analysis in Fig.~\ref{f:astroerr_snr_ifs_sd}, and for the KLIP analysis in Fig.~\ref{f:astroerr_snr_ifs_klip}.

We found that the trend of the standard deviation $\sigma$ could be described by the following formulas for the SD:

\begin{center}
\begin{equation}
    \sigma_{SD} =  0.25 + \frac{34.26}{S/N}~mas,       %0.254938      34.2582
\label{eq:astroifs_sd}
\end{equation}
\end{center}

\noindent for the model-independent analysis, plotted as a solid line in Fig.~\ref{f:astroerr_snr_ifs_sd}, and for the ideal case is calculated as:

\begin{center}
\begin{equation}
    \sigma_{SD} =  0.15 + \frac{15.64}{S/N}~mas,       %0.254938      34.2582
\label{eq:astroifs_sd}
\end{equation}
\end{center}  

\noindent  and plotted as a dashed line. Concerning KLIP analysis, the error bar is given by:

\begin{center}
\begin{equation}
    \sigma_{KLIP} = 0.21 + \frac{28.94}{S/N}~mas     %0.2158630      28.9392
\label{eq:astroifs}
\end{equation}
\end{center}

\noindent  for the model-independent analysis (\texttt{MPFIT2DPEAK}), plotted as a solid line in Fig.~\ref{f:astroerr_snr_ifs_klip}, and for the method of the ``negative synthetic planets'' with the same PSF is calculated as:

\begin{center}
\begin{equation}
    \sigma_{KLIP} = 0.05 + \frac{12.25}{S/N}~mas,     %0.2158630      28.9392
\label{eq:astroifs}
\end{equation}
\end{center}  

\noindent and plotted as a dashed line.

We consider that trends represent the error on our relative astrometric measurements with IFS. 

We can then predict that for a faint planet, with a contrast of $10^{-5}$, the typical relative position error will be of the order of 3~mas at a separation of 0\farcs35 from the host star, while for a brighter planet with a contrast of 3$\times10^{-4}$ will be of the order of 0.6~mas at the same separation. As discussed in the next Section, these results are comparable and even better to those obtained nowadays exploiting the ADI, even if they are obtained with only the use of SDI techniques.

\section{Conclusions}
\label{sec:conclusions}

%While SPHERE is ready to start its commissioning phase at VLT in Paranal, we can  of its forthcoming performance in characterizing direct imaged exoplanets within the solar neighborhood.  

During the AIT phase of SPHERE, the new planet finder for the VLT, we had the opportunity of testing the instrument, acquiring data, and analyzing them, exploring the capabilities of the IRDIFS and IRDIFS\_EXT modes. Using laboratory data and injecting synthetic planets into them, we were able to study the expected performance of these scientific IRDIFS modules IRDIS and IFS, when working in parallel (IRDIS in $H2H3$ and IFS in $YJ$). 

We performed the reduction of laboratory data using the SDI method for IRDIS, and the SD and KLIP methods for IFS. All of these methods provide good results in reducing the residual light from the primary star and detecting possible planetary candidates buried into the speckles pattern. As FoV rotation could not be simulated properly in the laboratory, our analysis does not include expected results using the ADI technique. Nonetheless, the 5-$\sigma$ contrast limit that we can reach using only spectral diversity methods is around 14~mag for IRDIS at separation of 0\farcs5 from the host star, and 15~mag for IFS.

To study the photometric and astrometric accuracy, we injected 750 synthetic planets in the same set of raw data acquired in the laboratory in parallel with IRDIS and IFS. We calculated the expected errors on photometry, astrometry, and spectroscopy in function of the S/N of the recovered companions. When observing on the sky, together with these minimum expected errors, we will have to take into account the uncertainties related to calibrations (e.g., for astrometry the orientation of the platescale and the determination of the true north). In order to reach these theoretical values of the errors accurate calibrations are needed.   

Using IRDIS we expect to detect a bright object like the white dwarf around the star HD\,8049, which has a $H$ band contrast of $10^{-2.8}$ and a projected separation of 1\farcs56 \citep{2013A&A...554A..21Z}, with a S/N of $\sim$270, an error on photometry of 0.045~mag and an error on the relative position of 0.2~mas. Using VLT/NACO, and exploiting the ADI, we have obtained errors of 0.12~mag on photometry and 7--10~mas on absolute astrometry.

The planet around $\beta$\,Pic is a 12 Myr old planet of $\sim$10~\MJup and semi-major axis of 9--10 AU \citep{2013A&A...555A.107B}. If it would be detected with IRDIS, for a $H$ band contrast of $10^{-4}$ and projected separation of 0\farcs5, the S/N would be of $\sim$22, the error on the photometry 0.16~mag and on the relative astrometric position the error would be of 1.1~mas. With VLT/NACO, typical errors are of the order of 0.2~mag for photometry and $\sim$13~mas for astrometry.

An error of 0.16~mag implies an uncertainty on the determination of the mass of the object of $\sim$0.5~\MJup using COND or DUSTY models \citep{2000ASPC..212..127A}, or of the order of 1--2~\MJup using core accretion models \citep{2008ApJ...683.1104F}. With IFS we would be able to retrieve a spectrum with error bars of the order of 0.15~mag on each channel and astrometric relative position error of 0.6~mas.

For fainter objects, such as the planets around the star HR\,8799 \citep{2008Sci...322.1348M}, SPHERE would be able to detect HR\,8799d, 7~\MJup planet at a distance of 27~AU \citep{2010Natur.468.1080M, 2013ApJ...768...24O}, with a relative astrometric error of 3~mas. \citet{2013A&A...549A..52E}, using LBT/PISCES, found an astrometric error of 10~mas. 

All these results will be revisited, once we are observing on the sky, when we will also be able to take advantage of ADI techniques. In particular, we expect that the KLIP reduction will greatly improve with the FoV rotation because the condition of the signal-less library will be better satisfied \citep[as described in][]{2014arXiv1409.6388P}.

The possibility of having a precise photometry and astrometry will contribute to the rejection of false alarms and to the characterization of the candidates and their host stars. Photometry and spectrometry will help to determine the temperature, the spectral type (that could possibly exclude background stars with flat spectrum) and the chemical composition of the atmosphere for bright targets.  Also, these data will make a great contribution to the study of the L-T transition \citep{2013ApJ...777...84B, 2013ApJ...768..121A}.

Astrometry with a precision of few mas will permit us to distinguish a background star from a comoving object over short temporal baselines for the follow-up. For example, the star $\beta$\,Pic has a proper motion of 83 mas/yr and the error bars on the position of the companion would be 0.7\% of the projected motion on the sky for one year. For HR\,8799, that has a proper motion of 118 mas/yr error bars will be the 2\% of the projected motion.

It is crucial to have high accuracy on the relative position of the planet to retrieve the orbital solution and discover possible perturbations due to other unseen planets. Astrometry could also give the possibility to calculate the mass of the companion by the motion of the primary with respect to the background object and it also opens the opportunity to determine the mass of stars in case of microlensing events \citep[see, e.g,][]{2014ApJ...782...89S} in very favorable cases.
 
\begin{acknowledgements}
We wish to thank the referee for her/his very fruitful comments. We are grateful with the SPHERE team and all the people always available during the tests at IPAG in Grenoble. The authors thank A.-L. Maire for useful discussions.  A.Z., D.M., and S.D. acknowledge partial support from PRIN INAF 2010 ``Planetary systems at young ages''. We acknowledge support from the 
French National Research Agency (ANR) through the GUEPARD project grant ANR10-BLANC0504-01. SPHERE is an instrument designed and built
by a consortium consisting of IPAG, MPIA, LAM, LESIA, Laboratoire Fizeau,
INAF, Observatoire de Gen\`eve, ETH, NOVA, ONERA, and ASTRON in collaboration with ESO.
\end{acknowledgements}

\bibliographystyle{aa}
\bibliography{irdifs}

\end{document}